\documentclass[english]{article}
\usepackage[T1]{fontenc}
\usepackage[latin9]{luainputenc}
\usepackage{geometry}
\usepackage{amsmath}
\usepackage{amsthm}
\usepackage{amssymb}
\usepackage{esint}
\usepackage{pgfplots}
\usepackage{subcaption}
\usetikzlibrary{fillbetween}
\usetikzlibrary{intersections}
\pgfplotsset{compat=newest}
\usepgfplotslibrary{fillbetween}
\usetikzlibrary{patterns}

\makeatletter

\newcommand*\LyXZeroWidthSpace{\hspace{0pt}}

\usepackage{beamerarticle,pgf}
\newcommand\makebeamertitle{\frame{\maketitle}}%
\AtBeginDocument{
	\let\origtableofcontents=\tableofcontents
	\def\tableofcontents{\@ifnextchar[{\origtableofcontents}{\gobbletableofcontents}}
	\def\gobbletableofcontents#1{\origtableofcontents}
}
\theoremstyle{definition}
\theoremstyle{plain}
\newtheorem{assumption}{\protect\assumptionname}
\theoremstyle{plain}
\newtheorem{prop}{\protect\propositionname}
\theoremstyle{plain}
\newtheorem{cor}{\protect\corollaryname}
\theoremstyle{definition}
\newtheorem{defn}{\protect\definitionname}
\theoremstyle{plain}
\newtheorem{lem}{\protect\lemmaname}
\theoremstyle{plain}
\newtheorem{thm}{\protect\theoremname}
\theoremstyle{remark}
\newtheorem{rem}{\protect\remarkname}

\makeatother

\usepackage{babel}
\providecommand{\assumptionname}{Assumption}
\providecommand{\corollaryname}{Corollary}
\providecommand{\definitionname}{Definition}

\providecommand{\lemmaname}{Lemma}
\providecommand{\propositionname}{Proposition}
\providecommand{\remarkname}{Remark}
\providecommand{\theoremname}{Theorem}

\begin{document}
\title{Identification with Posterior-Separable Latent Information Costs}
\author{Martín Bustos\thanks{University of Western Ontario. Email: mbustos3@uwo.ca}}
\makebeamertitle
\begin{abstract}
I provide a model of rational inattention with heterogeneity and prove
it is observationally equivalent to a state-dependent stochastic choice
model subject to attention costs. I demonstrate that additive separability
of unobservable heterogeneity, together with an independence assumption,
suffice for the empirical model to admit a representative agent. Using
conditional probabilities, I show how to identify: how covariates
affect the desirability of goods, (a measure of) welfare, factual
changes in welfare, and bounds on counterfactual market shares. 
\end{abstract}

\section{Introduction}
Seemingly mistaken decisions are sometimes rational. Often times,
choice situations involve uncertainty about the payoffs to different
alternatives. In principle individuals can gather information to learn
about outcomes. Regardless of the abundance of available information,
the extent to which uncertainty can be reduced is constrained when
information acquisition is costly. Information costs can arise due
to a variety of reasons, such as the cognitive effort to pay attention
and the opportunity cost of the time needed to attend to the environment.
Rationally inattention (RI) models incorporate attention as a scarce
resource. There is a trade-off between accuracy of information and
attention costs: on the one hand, the decision maker (DM) wants to
learn as much as she can so as to make better informed deicisons;
on the other hand, learning is costly in terms of attention effort.
Rationally inattentive behavior involves a two-step optimization problem:
first, the individual decides how to allocate her attention (i.e.
how much and what to learn about the true state of the world -the
true payoffs); then, she decides her choice given what was learned.

Recent evidence suggests attention costs play a key role in rejecting
the standard random utility model at describing aggregate behavior
(Aguiar, Boccardi, Kashaev, and Kim, 2023). Moreover, RI models have
been shown to have empirical content (Mat\v{e}jka and McKay, 2015;
Caplin, Dean, and Leahy, 2019) and nonparametric test for rationally
inattentive behavior have been provided (Caplin and Dean, 2015; Caplin,
Dean, and Leahy, 2022). Recent papers have proposed ways to estimate
demand that account for costly information acquisition when the characteristics
of the different alternatives are complex (Brown and Jeon, 2020) or
not directly observable (Joo, 2023), reaching to conclusions, such
that limiting the number of options available increases welfare, that
contradict standard discrete-choice models. In this paper, I aim at
contributing to bridge the gap between the theory of rational inattention
and empirical work.

Taking RI models to the data is challenging. First, most of the literature
is restricted to individual behavior, not accounting for heterogeneity.
Notable exceptions that include heterogeneity are recent papers proposing
ways to estimate these models (Brown and Jeon, 2023; Joo, 2023; Liao,
2024). Second, the standard cost function used in the literature,
the Shannon entropy, is too empirically restrictive. As shown in Mat\v{e}jka
and McKay (2015), in RI models with Shannon entropy, choice probabilities
follow a multinomial logit form. Albeit conveniently tractable, this
modeling choice comes at the cost of imposing the condition of independence
of irrelevant alternatives (IIA) on observable behavior, hence resulting
in unrealistic substitution patterns (i.e. the ratio of choice frequencies
between two alternatives does not change when adding or substracting
a third item from the menu). Recent efforts have been made to keep
tractability without harming empirical relevance. Caplin and Dean
(2015) use a generic cost function. Fosgerau, Melo, De Palma, and
Shum (2020) use Bregmann information costs, a generalized entropy
that allows for more realistic substitution patterns. Caplin et al.
(2022) propose two classes of posterior-separable cost functions that
allow for the use of standard Lagrangian methods to solve the model,
while relaxing the restrictions placed by the Shannon entropy. To
prevent undesirable behavioral restrictions, I present a theoretical
model that generalizes Caplin et al. (2022) by including additively
separable latent heterogeneity in both preferences and attention costs.
By introducing unobservable heterogeneity, I am able to describe the
population of rationally inattentive DMs. To my best knowkledge, this
is the first paper using RI for identifying demand that does not make
use of an entropy cost.

Third, most features of RI models cannot be directly observable. For
instance, attention allocations are modeled either as information
structures (e.g. Caplin and Dean, 2015) or Bayes-consistent mixtures
of posterior distributions (Caplin et al., 2022). For this reason,
the literature has provided an equivalent empirical counterpart: a
state-dependent stochastic choice (SDSC) model subject to information
costs (Mat\v{e}jka and McKay, 2015; Caplin et al., 2019; Caplin et
al., 2022). SDSC data consists of choice frequencies at each possible
(discrete) state of the world. I present a constrained SDSC model
with additively separable heterogeneity and prove it is equivalent
to my theoretical model. 

Fourth, most papers in the literature assume that the utilities are
known. Moreover, in these models, it is usually not possible to separately
identify preferences and attention costs. I address this issue by
adding observable covariates (i.e. alternative-specific attributes)
in the alternative-specific utility indices. Due to additive separability
of latent heterogeneity and an independence assumption, I show that
my model admits a representative agent and belongs to the class of
perturbed utility models (PUMs), thus allowing me to use the identification
results in Allen and Rehbeck (2019). Using conditional probabilities
of choice given observed covariates and states, I recover utility
indices, (a measure of) welfare, factual changes in welfare, and counterfactual
bounds on choice probabilities. 

Furthermore, by assuming that alternative-specific utility indices
are additively separable in states, I am able to extend the identification
results to stochastic choice data, that is, assuming the econometrician
observes conditional probabilities of choice given observed covariates
only. This latter result is of interest to practitioners of demand
estimation since the requirement on observables is only market-level
data.

More generally, my paper is related to models of costly information
acquisition. A strand of this literature studies models of motivated
cognition, where decision makers form beliefs and make choices under
uncertainty in the presence of a trade-off between accuracy and \textit{desirability}
of information. Individuals with motivated cognition typically either
have preferences over beliefs or derive anticipatory utility from
the flow of expected returns that investment in beliefs yield (Bénabou
and Tirole, 2016). Within the theory of motivated cognition, models
of wishful thinking predict that individuals manipulate their beliefs
to maximize their subjective expected utility net of belief-distorsion
costs (Caplin and Leahy, 2019; Kovach, 2020). The types of behavior
explained by wishful thinking span procrastination, confirmation bias,
and polarization. One of the most relevant applications of these models
is the market for assets as the behavior predicted by the theory explains
the occurrence of price bubbles.

The RI model with heterogeneity I provide is of general interest since
it might be extended to cover models of costly information acquisition
other than rational inattention, such as wishful thinking. Moreover,
it could be used to develop a statistical test for the null hypothesis
that a dataset of conditional distributions was generated by a population
of rationally inattentive consumers.\\

The rest of the paper proceeds as follows. Section 2 introduces my
RI model. Section 3 shows equivalence between this model and an empirical
counterpart. Section 4 covers the properties of the latter, which
are exploited for establishing identification with conditional mean
state-dependent stochastic choice data in Section 5. Section 6 imposes
additional structure to the model that enables identification with
conditional mean stochastic choice data. Section 7 concludes and discusses
future research.

\section{Model}
Consider a population of decision makers (DMs) who choose among a
finite set of alternatives $A$ whose payoffs vary with the occurrence
of different states of the world. The set of conceivable states of
the world $\Omega$ is finite and known to them, and at the moment
of making a decision, it is uncertain what the actual realization
of the state is. $\gamma\in\Delta(\Omega)$ is a belief about the
true state, where $\Delta(\Omega)$ denotes the set of probability
distributions on $\Omega$. DMs want to pick the alternative that
gives the highest perturbed expected utility,

\[
\gamma\cdot u_{a}(\omega)+\mathbf{E}(a),
\]
where $u_{a}(\omega)$ is a $|\Omega|$-dimensional vector whose $j^{th}$
component is the utility index $u_{a}(\omega_{j})$ representing the
desirability of item $a$ in state $\omega_{j}$, and the disturbance
function $\mathbf{E}:A\mapsto\mathbb{R}\cup\{-\infty\}$ denotes unobservable
heterogeneity in preferences, independent of the state.\footnote{Throughout the paper, I use $\textbf{boldface}$ to refer to random
objects.} The distrubance function can be interpreted as heterogeneity across
individuals in the population and across choice instances, i.e. a
preference shock. Importantly, at the moment of decision-making, the
realization of this disturbance function is known to the individual.

In this environment, DMs are endowed with a prior belief $\mu\in\Delta(\Omega)$,
which they can update by gathering information about the state. A
DM is said to learn something when the posterior belief $\gamma$
she forms differs from the prior. Learning, however, comes at a cost.
Let $\mathbf{T}\in\mathcal{T}$ be the posterior-specific attention
cost function, where $\mathcal{T}$ is the space of convex functions
$T:\Delta(\Omega)\mapsto\mathbb{R}_{+}\cup\{+\infty\}$ that satisfy
$T(\mu)=0$. Intuitively, learning nothing is costless and learning
something is weakly costly. This function represents latent heterogeneity
in disutility from attention effort. One may think of these attention
costs as cognitive effort from paying attention or the opportunity
cost of the time spent on learning. The realization of the attention
costs is also known to the individual at the moment of deciding. Define
\[
N^{a}(\gamma;\omega,\mathbf{E},\mathbf{T}):=\gamma\cdot u_{a}(\omega)+\mathbf{E}(a)-\mathbf{T}(\gamma),
\]
the net expected utility of choosing $a$ at $\gamma$. Note the trade-off
between accuracy of information and attention effort. Whereas in principle,
individuals would like to learn as much as possible about the true
state to make better informed decisions that lead to higher payoffs,
acquiring information is costly.

DMs face a two-step decision problem. The first step is deciding the
attention allocation: how much to learn from the environment. An attention
allocation $Q\in\Delta\left(\Delta(\Omega)\right)$ is a probability
distribution over beliefs, where $Q(\gamma)$ denotes the probability
of a posterior $\gamma$. Intuitively, the extent to which the formed
posterior belief $\gamma$ differs from the prior $\mu$ indicates
how much is learned. Let 

\[
\mathcal{Q}:=\left\{ Q\in\Delta\left(\Delta(\Omega)\right)\Big|\sum_{\gamma\in Supp(Q)}Q(\gamma)\gamma=\mu\right\} 
\]
be the set of feasible attention allocations, that is, the set of
distributions in $\Delta\left(\Delta(\Omega)\right)$ that satisfy
Bayes' rule.\footnote{The notation $SuppQ$ refers to the support of distribution $Q$.}
The second step is selecting a stochastic choice function $q:Supp(Q)\mapsto\Delta(A)$
which, for each realized posterior, gives a probability distribution
over alternatives in the choice set. Let 
\[
\Lambda:=\left\{ Q\in\mathcal{Q},q:Supp(Q)\mapsto\Delta(A)\right\} 
\]
be the set of feasible posterior-based policies. 

The perturbed expected utility of a given feasible posterior-based
policy $(Q,q)\in\Lambda$ is given by

\[
\sum_{\gamma\in Supp(Q)}\sum_{a\in A}Q(\gamma)q(a|\gamma)\left(\gamma\cdot u_{a}(\omega)+\mathbf{E}(a)\right).
\]
The expression

\[
\sum_{\gamma\in Supp(Q)}Q(\gamma)\mathbf{T}(\gamma)
\]
denotes the attention cost at a Bayes-consistent attention allocation
$Q\in\mathcal{Q}$.

A DM is said to be rationally inattentive if she chooses a pair $(\mathbf{Q},\mathbf{q})$
in the subset of feasible posterior-based policies $\boldsymbol{\Lambda}\subset\Lambda$
that majorize the perturbed expected utility net of attention costs,

\begin{equation}
\sup_{(Q,q)\in\Lambda}\sum_{\gamma\in Supp(Q)}\sum_{a\in A}Q(\gamma)q(a|\gamma)\left(\gamma\cdot u_{a}(\omega)+\mathbf{E}(a)-\mathbf{T}(\gamma)\right).\label{eq:1}
\end{equation}

Write (\ref{eq:1}) as a two-step optimization problem:

\[
\sup_{Q\in\mathcal{Q}}\sum_{\gamma\in Supp(Q)}Q(\gamma)\sup_{q(\gamma)\in\Delta(A)}\sum_{a\in A}q(a|\gamma)N^{a}(\gamma;\omega,\mathbf{E},\mathbf{T}).
\]
Fix any $\gamma\in SuppQ$. Then the second-stage problem satisfies:

\[
\sup_{q(\gamma)\in\Delta(A)}\sum_{a\in A}q(a|\gamma)N^{a}(\gamma,\omega,\mathbf{E},\mathbf{T})=N^{a}(\gamma;\omega,\mathbf{E},\mathbf{T}),
\]
for each $a\in A$ with $\mathbf{q}(a|\gamma)>0$. Define the value
of a posterior $\gamma$ as the maximized net expected utility at
$\gamma$,

\[
N(\gamma;\omega,\mathbf{E},\mathbf{T}):=\max_{a\in A}\left\{ N^{a}(\gamma;\omega,\mathbf{E},\mathbf{T})\right\} .
\]
Note that the maximized value attained in $(1)$ can then be written
as the optimization of posterior-specific values over Bayes-consistent
attention policies,
\begin{equation}
\sup_{Q\in\mathcal{Q}}\sum_{\gamma\in Supp(Q)}Q(\gamma)N(\gamma;\omega,\mathbf{E},\mathbf{T}).\label{eq:2}
\end{equation}

Before stating my first assumption, I introduce some concepts that
are necessary to characterize the solution to the model. Fix any $(E,T)$
in the support. In the interest of compactness in the notation, below
I will not explicitly write $(E,T)$ in the functions parametrized
by this tuple (e.g. $N(\gamma)\equiv N(\gamma;\omega,E,T)$). Define
the supporting function of the hypograph of $N$ in the direction
$\lambda\in\mathbb{R}^{|\Omega|}$,
\[
\delta^{*}(\lambda|\text{hyp}N):=\sup_{\gamma\in\Delta(\Omega),r\leq N(\gamma)}\lambda_{1}r+\lambda_{2}\cdot\gamma,
\]
and denote by $\varGamma(\lambda)$ the set of posteriors supported
by the supporting hyperplane in the direction $\lambda$,

\[
\varGamma(\lambda):=\left\{ \gamma\in\Delta(\Omega)\Big|\lambda_{1}N(\gamma)-\lambda_{2}\cdot\gamma=\delta^{*}(\lambda|\text{hyp}N)\right\} ,
\]
which generates the matrix

\[
\Gamma_{\lambda}:=[\gamma^{(1)}\;...\;\gamma^{(N)}],\;\text{where }\left\{ \gamma^{(1)}\;...\;\gamma^{(N)}\right\} =\varGamma(\lambda),
\]
with arbitrary order $n=1,...,N$.\footnote{See Section 13 in Rockafellar (1970) for a more thorough explanation
of the support function.}
\begin{example}
Consider $A=\{a,b\}$, $\Omega=\{\omega_{1},\omega_{2}\}$, $u_{a}(\omega_{1})=10$,
$u_{a}(\omega_{2})=5$, $u_{b}(\omega_{1})=12$, $u_{b}(\omega_{2})=3$,
$E(a)=1$, $E(b)=2$, and $T(\gamma)=4(\gamma-\frac{1}{2})^{2}$.
In Figure 1, see the plots for the net expected utility functions
$N^{a}$, $N^{b}$, the objective function $N$, the hypograph of
the objective function $hypN$, and examples of supporting hyperplanes
of $hypN$.
\end{example}
\LyXZeroWidthSpace{}

\begin{figure}[htbp]
\centering

\begin{subfigure}{.5\textwidth}
  \centering
  \begin{tikzpicture}[scale=0.6]
    \begin{axis}[
      axis lines=middle,
      xlabel={$\gamma$},
      ylabel={$\mathbb{R}_{+}$},
      xmin=0, xmax=1.2,
      ymin=0, ymax=14,
      domain=0:1,
      samples=100,
      xtick={0,1},
      ytick=\empty,
      xticklabels={0,1},
      yticklabels={},
      xlabel style={at={(axis description cs:1.05,0)},anchor=north},
      ylabel style={at={(axis description cs:0,1.05)},anchor=south},
      clip=false,
    ]

    \addplot[blue] expression {10*\x + (1-\x)*5 + 1 - 4*((\x-0.5)^2)};
    \node at (axis cs:1, {10*1 + (1-1)*5 + 1 - 4*((1-0.5)^2)}) [anchor=west] {$N^{a}$};

    \addplot[red] expression {12*\x + (1-\x)*3 + 2 - 4*((\x-0.5)^2)};
    \node at (axis cs:1, {12*1 + (1-1)*3 + 2 - 4*((1-0.5)^2)}) [anchor=west] {$N^{b}$};

    \draw[dashed] (axis cs:1,0) -- (axis cs:1,14);
    \end{axis}
  \end{tikzpicture}
  \caption{Net expected utilities of alternatives $a$ and $b$}
\end{subfigure}%
\begin{subfigure}{.5\textwidth}
  \centering
  \begin{tikzpicture}[scale=0.6]
    \begin{axis}[
      axis lines=middle,
      xlabel={$\gamma$},
      ylabel={$\mathbb{R}_{+}$},
      xmin=0, xmax=1.2,
      ymin=0, ymax=14,
      domain=0:1,
      samples=100,
      xtick={0,1},
      ytick=\empty,
      xticklabels={0,1},
      yticklabels={},
      xlabel style={at={(axis description cs:1.05,0)},anchor=north},
      ylabel style={at={(axis description cs:0,1.05)},anchor=south},
      clip=false,
    ]

    \addplot[thick, green!50!black] expression {max(10*\x + (1-\x)*5 + 1 - 4*((\x-0.5)^2), 12*\x + (1-\x)*3 + 2 - 4*((\x-0.5)^2))};
    \node at (axis cs:1, {max(10*1 + (1-1)*5 + 1 - 4*((1-0.5)^2), 12*1 + (1-1)*3 + 2 - 4*((1-0.5)^2))}) [anchor=west] {$N(\gamma)$};

    \draw[dashed] (axis cs:1,0) -- (axis cs:1,14);
    \end{axis}
  \end{tikzpicture}
  \caption{Objective function}
\end{subfigure}

\begin{subfigure}{.5\textwidth}
  \centering
  \begin{tikzpicture}[scale=0.6]
    \begin{axis}[
      axis lines=middle,
      xlabel={$\gamma$},
      ylabel={$\mathbb{R}_{+}$},
      xmin=0, xmax=1.2,
      ymin=0, ymax=14,
      domain=0:1,
      samples=200,
      xtick={0,1},
      ytick=\empty,
      xticklabels={0,1},
      yticklabels={},
      xlabel style={at={(axis description cs:1.05,0)},anchor=north},
      ylabel style={at={(axis description cs:0,1.05)},anchor=south},
      clip=false,
    ]

    \addplot[green!50!black, fill=green!50!black, opacity=0.5, domain=0:1] {
      max(10*x + (1-x)*5 + 1 - 4*((x-0.5)^2), 12*x + (1-x)*3 + 2 - 4*((x-0.5)^2))
    } \closedcycle;

    \node[anchor=west] at (axis cs:0.5,{max(10*0.5 + (1-0.5)*5 + 1 - 4*((0.5-0.5)^2), 12*0.5 + (1-0.5)*3 + 2 - 4*((0.5-0.5)^2)) - 1}) {hyp$N$};

    \draw[dashed] (axis cs:1,0) -- (axis cs:1,14);
    \end{axis}
  \end{tikzpicture}
  \caption{Hypograph of the objective function}
\end{subfigure}%
\begin{subfigure}{.5\textwidth}
  \centering
  \begin{tikzpicture}[scale=0.6]
    \begin{axis}[
      axis lines=middle,
      xlabel={$\gamma$},
      ylabel={$\mathbb{R}_{+}$},
      xmin=0, xmax=1.2,
      ymin=0, ymax=14,
      domain=0:1,
      samples=100,
      xtick={0,1},
      ytick=\empty,
      xticklabels={0,1},
      yticklabels={},
      xlabel style={at={(axis description cs:1.05,0)},anchor=north},
      ylabel style={at={(axis description cs:0,1.05)},anchor=south},
      clip=false,
    ]

    \def\Na{10*x + (1-x)*5 + 1 - 4*((x-0.5)^2)}
    \def\Nb{12*x + (1-x)*3 + 2 - 4*((x-0.5)^2)}

    \addplot[green!50!black, fill=green!50!black, opacity=0.5, domain=0:1] {
      max(\Na, \Nb)
    } \closedcycle;
    \node[anchor=west] at (axis cs:0.5,3) {hyp$N$};

    \addplot[thick, color=red] expression {7*x + 25/4};

    \node[circle, fill, inner sep=2pt, color=red] at (axis cs:3/4, 23/2) {};

    \addplot[thick, color=red] expression {8*x + 89/16};

    \node[circle, fill, inner sep=2pt, color=red] at (axis cs:5/8, 169/16) {};

    \draw[dashed] (axis cs:1,0) -- (axis cs:1,14);
    \end{axis}
  \end{tikzpicture}
  \caption{Two supporting hyperplanes of the hypograph of the objective function}
\end{subfigure}

\caption{}
\end{figure}

\begin{assumption}
Suppose the following: $(i)$ $\mathbf{Q}$, $\mathbf{q}$, $\mathbf{E}$,
and $\mathbf{T}$ satisfy (\ref{eq:1}); $(ii)$ $\mathbf{E}(a)\neq-\infty$
for some $a\in A$, $\mathbf{T}$ is strictly convex in $\gamma$;
$(iii)$ the joint distribution of $(\mathbf{E},\mathbf{T})$ satisfies
\[
\mathbb{P}\left(\left\{ (E,T)\in\mathcal{E}\times\mathcal{T}\Big|\text{rank}\Gamma_{\lambda}(E,T)=|\varGamma(\lambda,E,T)|\right\} \right)=1,
\]
for all directions $\lambda\in\mathbb{R}^{|\Omega|}$; and $(iv)$
for any arbitrary order $j=1,...,J=|\Omega|$,
\[
(u_{a,\omega_{1}}-u_{a,\omega_{J}},...,u_{a,\omega_{J-1}}-u_{a,\omega_{J}})^{T}\ne(u_{b,\omega_{1}}-u_{a,\omega_{J}},...,u_{b,\omega_{J}}-u_{a,\omega_{J}})^{T}
\]
 for each $a\in A$, $b\in A\backslash\{b\}$.
\end{assumption}
In Assumption 1, condition $(i)$ means that the model in (\ref{eq:1})
is the data-generating process. Condition $(ii)$ ensures that the
disturbance function is such that there is always at least one alternative
$a$ whose perturbed expected utility is a real number. Since the
attention cost function evaluated at the prior is $0$, then the feasible
strategy $(Q,q)$ with $Q(\mu)=1$ and $q(a|\mu)=1$ gives always
a real value. Therefore, the model has at least one maximizer. Condition
$(iii)$ implies that for any direction, the number of posteriors
on the supporting hyperplane is at most as large as the number of
possible states. This condition prevents the model from having mutiple
optimal attention allocations. Finally, condition $(iv)$, imposes
that for each pair of alternatives $a$ and $b$, their corresponding
net expected utilities $N^{a}(\gamma)$ and $N^{b}(\gamma)$ have
different slopes at each posterior belief $\gamma$. This condition
precludes optimal stochastic choice functions from giving nondegenerate
distributions at any posterior supported by the optimal attention
allocation. In other words, indifference among alternatives is ruled
out at each optimal posterior, thus resulting in a unique optimal
choice function. Overall, the conditions stated in this assumption
suffice for an optimizer of the model to exist and to be unique.

\LyXZeroWidthSpace{}
\begin{prop}
(Uniqueness). Let Assumption 1 hold. Then for any fixed $(D,T)$ in
the support, $(\hat{Q},\hat{q})\in\Lambda$ that solves (\ref{eq:1})
exists and is unique. In particular, the optimal attention policy
$\hat{Q}$ is the fully-supported distribution over posteriors supported
by the hyperplane tangent to the lower concave envelope of the function
$N$ at the prior $\mu$, and the optimal choice $\hat{q}$ maps each
optimal posterior $\hat{\gamma}\in Supp(\hat{Q})$ to a degenerate
distribution on $A$.
\end{prop}
\LyXZeroWidthSpace{}

For any $(D,T)$ in the support, an optimal posterior $\hat{\gamma}$
is said to be uniquely associated with an alternative $a$ in the
choice set if at $\hat{\gamma}$, $a$ is chosen with probability
$1$ and $a$ is not chosen with positive probability at any other
optimal posterior.

Due to functions $\left(N^{a}\right){}_{a\in A}$ being strictly concave
in $\gamma$ and since at each optimal posterior, there are no ties
among alternatives, then it follows that the supporting hyperplane
that characterizes optimal posteriors can be tangent to any function
$N^{a}$ at most at one point. As a result, each optimal posterior
is uniquely associated with one alternative in the choice set, i.e.
if $(\hat{Q},\hat{q})$ is optimal, then it follows that for each
$\hat{\gamma}\in Supp(\hat{Q})$ there is $a\in A$ such that $\hat{q}(a|\hat{\gamma})=1$
and $\hat{q}(a|\hat{\gamma}')=0$ for each $\hat{\gamma}'\in Supp(\hat{Q})\backslash\{\hat{\gamma}\}$.
I formalize this result below.

Define the set of posteriors at which, the optimal choice is $a$,

\[
\gamma^{a}:=\left\{ \gamma\in Supp(Q)|q(a|\gamma)=1\right\} .
\]

\begin{cor}
Under Assumption 1, for any fixed $(E,T)$, the solution $(\hat{Q},\hat{q})$
to $(1)$ is such that each $\hat{\gamma}\in Supp(\hat{Q})$ is such
that $\gamma^{a}=\{\hat{\gamma}\}$ for a unique $a\in A$. 

\LyXZeroWidthSpace{}
\end{cor}
If an optimal posterior $\hat{\gamma}$ is uniquely associated with
an alternative $a$ in the choice set, it is known that when $\hat{\gamma}$
is realized, $a$ is picked with probability $1$, and that the only
optimal posterior at which $a$ is chosen is $\hat{\gamma}$. This
property will be crucial to map the model to observables in the next
section.

Finally, note that (\ref{eq:1}) can be further reexpressed as

\begin{equation}
\sup_{Q\in\mathcal{Q}}\sum_{\{a\in A|\exists\gamma\in SuppQ:\gamma=\boldsymbol{\gamma}^{a}\}}Q(\boldsymbol{\gamma}^{a})N^{a}(\boldsymbol{\gamma}^{a};\omega,\mathbf{E},\mathbf{T}).\label{eq:3}
\end{equation}

\section{State-Dependent Stochastic Choice}
In this section, I explore the link between my model with heterogeneity
and state-dependent stochastic choice data, the empirical primitive
in the literature. Such datasets comprise choice frequencies at each
possible state. 

Define $\mathcal{P}:=\{P:\Omega\mapsto\Delta(A)\}$, the set of state-dependent
stochastic choice (SDSC) functions. While attention allocations and
choice functions are not directly observable, SDSC functions might
be in principle observed by the analyst. I then study how, according
to my model, decisions by rationally inattentive individuals relate
to SDSC data.

\LyXZeroWidthSpace{} 
\begin{defn}
(Generated State-Dependent Stochastic Choice Data) Let $P\in\mathcal{P}$
be the SDSC data generated by $(Q,q)\in\Lambda$, where for each $\omega\in\Omega$,
$a\in A$,

\[
P(a|\omega):=\mu(\omega)^{-1}\sum_{\boldsymbol{\gamma}\in Supp(Q)}Q(\gamma)q(a|\gamma)\gamma(\omega),
\]
is the conditional probability of picking item $a$ at state $\omega$,
and for each $a\in A$,

\[
P(a):=\sum_{\boldsymbol{\gamma}\in Supp(Q)}Q(\gamma)q(a|\gamma)
\]
denotes the unconditional probability of choosing $a$. 
\end{defn}
Uniqueness of the maximizer in the rational inattention model entails
that, for any given realization of heterogeneity $(D,T)$, the pair
consisting of the optimal attention allocation and the optimal choice
function $(Q,q)$ unequivocally induces a unique SDSC function $P$.
I formalize this result below.
\begin{lem}
Let Assumption 1 hold. Then the SDSC function $\mathbf{P}$ generated
by $(\mathbf{Q},\mathbf{q})$ that solves (\ref{eq:1}) is unique
and such that for each $a\in A$, it satisfies

\[
\mathbf{P}(a|\omega)=\begin{cases}
\mu(\omega)^{-1}\mathbf{Q}(\gamma)\boldsymbol{\gamma}^{a}(\omega), & \text{if }\boldsymbol{\gamma}^{a}\in Supp(\mathbf{Q})\\
0, & \text{otherwise},
\end{cases}
\]
for each $\omega\in\Omega$, and consequently,

\[
\mathbf{P}(a)=\begin{cases}
\mathbf{Q}(\boldsymbol{\gamma}^{a}), & \text{if }\boldsymbol{\gamma}^{a}\in Supp(\mathbf{Q})\\
0 & \text{otherwise}.
\end{cases}
\]
\end{lem}
\begin{proof}
By Proposition 1, $(\mathbf{Q},\mathbf{q})$ solving (\ref{eq:1})
is unique and such that at each $\boldsymbol{\gamma}\in Supp(\mathbf{Q})$,
$\mathbf{q}(\boldsymbol{\gamma})$ is a degenerate distribution over
$A$. Moreover, by Corollary 1, each $\boldsymbol{\gamma}\in Supp(\mathbf{Q})$
satisfies $\boldsymbol{\gamma}=\boldsymbol{\gamma}^{a}$ for some
$a\in A$. 
\end{proof}
\begin{cor}
Under Assumption 1, the optimal posterior uniquely associated with
$a$ is such that satisfies

\[
\boldsymbol{\gamma}^{a}(\omega)=\frac{\mathbf{P}(a|\omega)\mu(\omega)}{\mathbf{P}(a)},\;\text{for each }\omega\in\Omega.
\]
\end{cor}
\LyXZeroWidthSpace{}

Next, I show what plausibly observable SDSC data reveals from rationally
inattentive behavior and provide a state-dependent stochastic choice
model subject to attention costs with latent heterogeneity.

\LyXZeroWidthSpace{}
\begin{defn}
Define the following objects revealed by a SDSC function $\rho\in\mathcal{P}$:
(Revealed posteriors) $\bar{\gamma}^{a}(\omega):=\frac{\rho(a|\omega)\mu(\omega)}{\rho(a)}$,
for each $a\in Supp(\rho)$, $\omega\in\Omega$; (Revealed attention)
$\bar{Q}(\gamma):=\sum_{\{a\in A|\gamma=\bar{\gamma}^{a}\}}\rho(a)$,
for each $\gamma\in\Delta(\Omega)$; and (Revealed choice) $\bar{q}(a|\bar{\gamma}^{a}):=\frac{\rho(a)}{\bar{Q}(\bar{\gamma}^{a})}$,
for each $a\in Supp(\rho)$ and $0$ otherwise.
\end{defn}
Now consider an alternative model where the choice set, uncertainty
about payoffs, prior beliefs, and latent heterogeneity, all remain
the same as in (\ref{eq:1}), but where, instead of solving a two-step
problem, DMs make a single-step decision. In other words, given a
choice set $A$ and an endowment of utility indices $u$ and prior
belief $\mu$, individuals drawing latent functions $\mathbf{E}$
and $\mathbf{T}$ choose a SDSC function $\rho\in\mathcal{P}$ that
gives a probability distribution over alternatives in $A$ at each
state $\omega\in\Omega$. In this model, the perturbed expected utility
of an item $a$ is

\[
\sum_{j=1}^{J}\frac{\rho(a|\omega_{j})\mu(\omega_{j})}{\rho(a)}u_{a}(\omega_{j})+\mathbf{E}(a).
\]
Let $f$ be a vector-valued function $f:\mathcal{P}\times A\mapsto\Delta(\Omega)$
where for $\rho\in\mathcal{P}$, $a\in A$, the $j^{th}$ entry of
$f(\rho;a)$ is defined as $f_{j}\left(\rho;a\right):=\frac{\rho(a|\omega_{j})\mu(\omega_{j})}{\rho(a)}$
for $a\in Supp(\rho)$, $\omega_{j}\in\Omega$, and $f\left(\rho;a\right)=0_{|\Omega|-1}$
for \textbf{$a\in A\backslash\{Supp(\rho)\}$}. Interpret the function
$f$ evaluated at $(\rho;a)$ as the posterior belief resulting from
$\rho(a)$. The net expected utility of $a$ at $f(\rho;a)$ is

\[
N^{a}\left(f(\rho;a);\omega,\mathbf{E},\mathbf{T}\right):=f(\rho;a)\cdot u_{a}(\omega)+\mathbf{E}(a)-\mathbf{T}\left(f(\rho;a)\right).
\]

$\boldsymbol{\rho}\in\mathcal{P}$ is chosen to solve

\begin{equation}
\sup_{\rho\in\mathcal{P}}\sum_{a\in A}\rho(a)\left[f(\rho;a)\cdot u_{a}(\omega)+\mathbf{E}(a)-\mathbf{T}\left(f(\rho;a)\right)\right].\label{eq:4}
\end{equation}

Note that by definition of $N^{a}$, the model can be equivalently
expressed as

\[
\sup_{\rho\in\mathcal{P}}\sum_{a\in A}\rho(a)N^{a}\left(f(\rho;a);\omega,\mathbf{E},\mathbf{T}\right).
\]
Notice further that under Assumption 1 $(iv)$, $a\in Supp(\rho)$
implies that for any $(E,T)$ in the support,

\[
N^{a}\left(f(\rho;a);\omega,E,T\right)>N^{b}\left(f(\rho;a);\omega,E,T\right),\;\text{for each }b\in A.
\]
Hence, by definition of $N$, the model can be further reexpressed
as

\begin{equation}
\sup_{\rho\in\mathcal{P}}\sum_{a\in A}\rho(a)N\left(f(\rho;a);\omega,\mathbf{E},\mathbf{T}\right).\label{eq:5}
\end{equation}

Under my assumptions, this model too has a unique maximizer. As a
result, fixing any latent functions $(E,T)$ in the support, revealed
posteriors, revealed attention and revealed choice functions may be
recovered from SDSC data.
\begin{prop}
Let Assumption 1 hold. Then there exists a unique $\boldsymbol{\rho}\in\mathcal{P}$
that solves (\ref{eq:4}). Moreover, for each $a\in Supp(\boldsymbol{\rho})$
the following conditions are met: the revealed posterior satisfies
$\bar{\boldsymbol{\gamma}}^{a}(\omega_{j})=\frac{\boldsymbol{\rho}(b|\omega_{j})\mu(\omega_{j})}{\boldsymbol{\rho}(b)}$
for each $\omega_{j}\in\Omega$ only if $b=a$; the revealed attention
satisfies $\bar{\mathbf{Q}}(\bar{\boldsymbol{\gamma}}^{a})=\boldsymbol{\rho}(a)$;
and the revealed choice is such that 
\[
\bar{\mathbf{q}}(b|\bar{\boldsymbol{\gamma}}^{a})=\begin{cases}
1, & \text{for }b=a\\
0, & \text{otherwise,}
\end{cases}\;\text{for each }b\in A.
\]
\end{prop}
Uniqueness in both models allows me to enunciate the main result of
the paper.
\begin{thm}
(Equivalence) Let Assumption 1 hold. Then, the following propositions
are equivalent:

$(i)$ The posterior-based policy $(\mathbf{Q},\mathbf{q})\in\Lambda$
is the maximizer of (\ref{eq:1});

$(ii)$ The SDSC function $\mathbf{P}\in\mathcal{P}$ generated by
$(\mathbf{Q},\mathbf{q})\in\Lambda$ is the maximizer of (\ref{eq:4}).

$(iii)$ The SDSC function $\boldsymbol{\rho}\in\mathcal{P}$ is the
maximizer of (\ref{eq:4}); and

$(iv)$ The pair consisting of an attention and choice distribution
$(\bar{\mathbf{Q}},\bar{\mathbf{q}})\in\Lambda$ revealed by $\boldsymbol{\rho}\in\mathcal{P}$
is the maximizer of (\ref{eq:1}).
\end{thm}
Under Assumption 1, the optimizer of each version of the model satisfies
the one-to-one mapping property with respect to the other. If a pair
of an attention allocation and a choice function is the optimizer
(\ref{eq:1}), then it follows that its empirical counterpart, its
generated SDSC function, is the maximizer of (\ref{eq:4}). On the
other hand, if a SDSC function is the maximizer (\ref{eq:4}), then
its theoretical counterpart, the pair consisting of its revealed attention
and revealed choice function, is the optimizer of (\ref{eq:1}).

One of the reasons why this result is critical to learn about aggregate
demand of a population of rationally inattentive individuals is because
it implies the theoretical model and its empirical counterpart, the
constrained SDSC model, can be used interchangeably. This is useful
as mean SDSC data might in principle be observable. 

\section{Properties of the SDSC model}
The SDSC model presented before is a perturbed utility model (PUM).
In the interest of identifying structural and counterfactual parameters
related to changes in prices and other attributes, I extend the model
to incorporate covariates. I then impose assumptions sufficient to
derive desirable aggregation properties of the model and features
of its optimization structure that enable identification with conditional
mean SDSC data. 

For each alternative $a\in A$, let $\mathbf{x}_{a}\in\mathcal{X}_{a}\subseteq\mathbb{R}^{L_{a}}$
be a random vector listing attributes of $a$. For each $a\in A$,
$\omega_{j}\in\Omega$, let $u_{a}(.,\omega_{j}):\mathcal{X}_{a}\mapsto\mathbb{R}$
be a utility index that represents how attributes affect the desirability
of item $a$ at state $\omega$. Denote $u_{a}(x_{a},\omega):=\left(u_{a}(x_{a},\omega_{1}),\;...,\;u_{a,}(x_{a},\omega_{J})\right)^{T}$,
the vector of utilities of item $a$ evaluated at $x_{a}$ at each
possible state $\omega_{1},...,\omega_{J}\in\Omega$.

\LyXZeroWidthSpace{}
\begin{assumption}
Suppose the following: $(i)$ $\mathbf{Q}$, $\mathbf{q}$, $\mathbf{x}$,
$\mathbf{E}$, and $\mathbf{T}$ satisfy

\[
\sup_{(Q,q)\in\Lambda}\sum_{\gamma\in Supp(Q)}\sum_{a\in A}Q(\gamma)q(a|\gamma)\left(\gamma\cdot u_{a}(\mathbf{x}_{a},\omega)+\mathbf{E}(a)-\mathbf{T}(\gamma)\right);
\]
and $(ii)$ $\mathbf{x}$ and $(\mathbf{E},\mathbf{T})$ are independent. 
\end{assumption}

\subsection{Aggregation}
For the purpose of identification of structural and counterfactual
parameters, the SDSC model is useful, not only due to being the observable
counterpart of the theoretical model, but also because it keeps the
additively separability property of latent heterogeneity. This condition
together with the independence assumption allow me to link my rational
inattention model with latent heterogeneity to the identification
framework in Allen and Rehbeck (2019). 

Concretely, additively separable latent heterogeneity and independence
are the key conditions that suffice for the model to admit a representative
agent, which is key for establishing identification under the assumption
that conditional mean SDSC data is observable.
\begin{lem}
(Theorem 1 in Allen and Rehbeck {[}2019{]}). Let Assumption 1 hold
except $(i)$. Let Assumption 2 hold. Let $x\in Supp(\mathbf{x})$
and suppose $\mathbf{P}$ is $(\mathbf{x},\mathbf{E},\mathbf{T})$-measurable,
and both $\mathbb{E}[\mathbf{P}|\mathbf{x}=x]$ and
\[
\mathbb{E}\left[\sum_{a\in A}\mathbf{P}(a)\left(\mathbf{E}(a)-\mathbf{T}(\boldsymbol{\gamma}^{a})\right)\right]
\]
exist and are finite, where the expectations are over the joint distribution
of $(\mathbf{E},\mathbf{T})$. Then it follows that $(i)$ $\mathbb{E}[\mathbf{P}|\mathbf{x}=x]$
is an optimizer of the representative agent problem,
\[
\sup_{P\in\mathcal{P}}\sum_{\omega\in\Omega}\sum_{a\in A}P(a|\omega)\mu(\omega)u_{a}(x_{a},\omega)+\bar{D}(P),
\]
with
\[
\bar{D}(P)=\sup_{\left\{ P'\in\mathcal{P}_{(\mathbf{E},\mathbf{T})}|\mathbb{E}[P']=P\right\} }\mathbb{E}\left[\sum_{a\in A}P'(a)\left(\mathbf{E}(a)-\mathbf{T}\left(f(P',a)\right)\right)\right],
\]
where $\mathcal{P}_{(\mathbf{E},\mathbf{T})}$ is the set of $(\mathbf{E},\mathbf{T})$-measurable
functions that map to $\mathcal{P}$. $(ii)$ Define the indirect
utility function of the representative agent problem,

\[
V(v):=\sup_{P\in\mathcal{P}}\sum_{\omega\in\Omega}\sum_{a\in A}P(a|\omega)\mu(\omega)v_{a,\omega}+\bar{D}(P).
\]
Thus, $V$ satisfies

\[
V\left(u(x,\omega)\right)=\mathbb{E}\left[\sup_{P\in\mathcal{P}}\sum_{\omega\in\Omega}\sum_{a\in A}P(a|\omega)\mu(\omega)u_{a}(x_{a},\omega)+\sum_{a\in A}P(a)\left(\mathbf{E}(a)-\mathbf{T}\left(f(P,a)\right)\right)\right].
\]
\end{lem}
\LyXZeroWidthSpace{}

This aggregation result renders identification of the distribution
of unobservable heterogeneity unnecessary to identify utility indices
or mean indirect utility, as well as specifies the data requirements
for identification. Specifically, it is sufficient to observe conditional
mean SDSC functions $\left(\mathbb{E}[\mathbf{P}|\omega,\mathbf{x}=x]\right)_{\omega\in\Omega}$. 
Note that in both the problem of the representative agent and the
average indirect utility $V$, optimization is over function in $\mathcal{P}$.
According to the original theorem in Allen and Rehbeck (2019), optimization
in both cases is over the convex hull of the set $\mathcal{P}$. However,
the convex hull of $\mathcal{P}$ is the set $\mathcal{P}$ itself.

As the aggregation theorem requires the convex hull of the feasibility
set, using the SDSC model is necessary for measure-theoretic reasons.
If one considers the theoretical model, the convex hull of $\Lambda$
is $\Lambda$ itself. It is clear that, in contrast to $\mathcal{P}$,
$\mathcal{Q}$ is an infinite-dimensional set, and hence, does not
fit in the theorem proposed by Allen and Rehbeck (2019).

\subsection{Structure}
Building upon the aggregation result, I am able to exploit the optimization
structure of the model for identification. Concretely, I apply an
envelope theorem to the representative agent model and leverage asymmetries
of cross-parital derivatives of the conditional mean SDSC function.

Below, I derive the properties that follow from the optimization structure
of the model and leave the identification results for the next section.
\begin{rem}
$\text{Conv}\mathcal{P}$ is a nonempty, closed and convex set.
\end{rem}
\begin{assumption}
Suppose the following: $(i)$ $\mathbb{E}[\mathbf{P}|\mathbf{x}=x]$
is the unique optimizer of

\[
\sup_{P\in\mathcal{P}}\underset{\omega\in\Omega}{\sum}\sum_{a\in A}P(a|\omega)\mu(\omega)u_{a}(x_{a},\omega)+\bar{D}(P),
\]
for each $x\in Supp(\mathbf{x})$; $(ii)$ $\bar{D}:\mathcal{P}\mapsto\mathbb{R\cup}\{-\infty\}$
is concave, upper semi--continuous, and finite at some $P\in\mathcal{P}$.
\end{assumption}
\begin{lem}
(Roy's identity {[}Lemma 2 in Allen and Rehbeck (2019){]} Let Assumption
3 hold. Then, for each $a\in A$, $\omega_{j}\in\Omega$,
\[
\mathbb{E}[\mathbf{P}(a|\omega_{j})|\mathbf{x}=x]\mu(\omega_{j})=\frac{\partial}{\partial v_{a,j}}V(\vec{v})\Bigg|_{v=u(x,\omega)}.
\]
\end{lem}
\begin{lem}
(Slutsky conditions {[}Lemma 2 in Allen and Rehbeck (2019){]}). Let
Assumption 3 hold and let $V$ be twice continuously differentiable
in a neighbourhood of $v\in\mathbb{R}^{|A|\times|\Omega|}$. Then,
the following conditions are satisfied: $(i)$ (Positive semi-definitiness)
$\lambda^{T}\nabla^{2}V(v)\lambda\geq0$, for all $\lambda\in\mathbb{R}^{|A|\times|\Omega|}$;
and $(ii)$ (Symmetry) $\frac{\partial^{2}}{\partial v_{a,j}\partial v_{b,j}}V(v)=\frac{\partial^{2}}{\partial v_{b,j}\partial v_{a,j}}V(v)$
for all $a,b\in A$ and $j=1,...,J$.
\end{lem}
\begin{lem}
{[}Lemma 3 in Allen and Rehbeck (2019){]} Let Assumption 3 hold. Define
\[
\bar{D}_{\text{\ensuremath{\mathcal{P}}}}(P):=\begin{cases}
\bar{D}(P), & \text{if }P\in\mathcal{P}\\
-\infty, & \text{otherwise}.
\end{cases}
\]
If $\bar{D}_{\text{\ensuremath{\mathcal{P}}}}(P)$ is differentiable
at $\mathbb{E}[\mathbf{P}|\mathbf{x}=x]$, then for each $a\in A$,
$\omega_{j}\in\Omega$,

\[
\mu(\omega_{j})u_{a}(x_{a},\omega_{j})=-\frac{\partial}{\partial P(a|\omega_{j})}\bar{D}_{\text{\ensuremath{\mathcal{P}}}}(P)\Bigg|_{P=\mathbb{E}[\mathbf{P}|\mathbf{x}=x]}.
\]
\end{lem}

\section{Identification with SDSC data}

The fact that the model admits a representative agent indicates that
observing the conditional mean SDSC function is the only data requirement
for identification. This is to say, it is enough for the analyst to
observe the conditional probability distribution of choice at each
possible state. 
\begin{assumption}
Suppose the following: $(i)$ (Observables) the econometrician observes
(can consistently estimate) the mean generated SDSC data

\[
\left\{ \mathbb{P}(a|\omega_{j},\mathbf{x}=x)\right\} _{a\in A,\omega_{j}\in\Omega},
\]
for each $x\in Supp(\mathbf{x})$, where $\mathbb{P}(a|\omega_{j},\mathbf{x}=x)=\mathbb{E}[\mathbf{P}(a|\omega_{j})|\mathbf{x}=x]$
and the expectation is over the joint distribution of $(\mathbf{E},\mathbf{T})$;
and $(ii)$ (Full support) for each $\omega_{j}\in\Omega$, $x\in Supp(\mathbf{x})$,
$\mathbb{E}[\mathbf{P}(a|\omega_{j})|\mathbf{x}=x]>0$ for each $a\in A$.
\end{assumption}

\subsection{Identification of good-state-specific utility indices}

Using assymmetry of the cross-partial derivatives of the conditional
mean SDSC function and an exclusion restriction (i.e. covariates are
exclusive to one alternative), I identify the utility indices up to
location and scale without having to recover the distribution of latent
heterogeneity.
\begin{defn}
$\frac{\partial}{\partial x_{a,p}}u_{a}(x_{a},\omega_{j})|_{x_{a}=x_{a}^{*}}$
and $\frac{\partial}{\partial x_{b,q}}u_{b}(x_{b},\omega_{j})|_{x_{b}=x_{b}^{*}}$
are said to be paired if they exist and the following conditions holds:
$(i)$ There exists a known value $x^{*}\in Supp(\mathbf{x})$ whose
$a^{th}$ and $b^{th}$ components are $x_{a}^{*}$ and $x_{b}^{*}$,
respectively; $(ii)$ $\frac{\partial}{\partial x_{a,p}}\mathbb{E}[\mathbf{P}(a|\omega_{j})|\mathbf{x}=x]|_{x=x^{*}}$
and $\frac{\partial}{\partial x_{b,q}}\mathbb{E}[\mathbf{P}(b|\omega_{j})|\mathbf{x}=x]|_{x=x^{*}}$
exist; $(iii)$ $V$ is twice continuously differentiable in a neighbourhood
of $u(x^{*})$; and $\frac{\partial^{2}}{\partial v_{a,j}\partial v_{b,j}}V(v)|_{v=u(x^{*},\omega)}\neq0$.
If, in addition, $\frac{\partial}{\partial x_{a,p}}u_{a}(x_{a},\omega_{j})|_{x_{a}=x_{a}^{*}}\neq0$
and $\frac{\partial}{\partial x_{b,q}}u_{b}(x_{b},\omega_{j})|_{x_{b}=x_{b}^{*}}\neq0$,
then they are said to be strictly paired.
\end{defn}
\begin{lem}
{[}Proposition 1 in Allen and Rehbeck (2019){]} Let Assumption 3 hold
and assume $x_{a,p}$ and $x_{b,q}$ are regresssors specific to $a$
and $b$, respectively. If the points $\frac{\partial}{\partial x_{a,p}}u_{a}(x_{a},\omega_{j})|_{x_{a}=x_{a}^{*}}$
and $\frac{\partial}{\partial x_{b,q}}u_{b}(x_{b},\omega_{j})|_{x_{b}=x_{b}^{*}}$
are paired and $\frac{\partial}{\partial x_{b,p}}u_{b}(x_{b},\omega_{j})|_{x_{b}=x_{b}^{*}}\neq0$,
then there is some known $x^{*}\in Supp(\mathbf{x})$ such that
\begin{equation}
\dfrac{\frac{\partial}{\partial x_{a,p}}\mathbb{E}[\mathbf{P}(b|\omega_{j})|\mathbf{x}=x]\Big|_{x=x^{*}}}{\frac{\partial}{\partial x_{b,q}}\mathbb{E}[\mathbf{P}(a|\omega_{j})|\mathbf{x}=x]\Big|_{x=x^{*}}}=\dfrac{\frac{\partial}{\partial x_{a,p}}u_{a}(x_{a},\omega_{j})\Big|_{x_{a}=x_{a}^{*}}}{\frac{\partial}{\partial x_{b,q}}u_{b}(x_{b},\omega_{j})\Big|_{x_{b}=x_{b}^{*}}}.\label{eq:6}
\end{equation}
Suppose further that Assumption 4 holds, then the right-hand side
of $(3)$ is identified.
\end{lem}
\begin{lem}
{[}Proposition 2 in Allen and Rehbeck (2019){]} Let Assumptions 3
and 4 hold. Assume that all regressors are good-specific and $|A|,|\Omega|\geq2$.
Assume $\mathbf{x}$ has a rectangular support; $u$ is differentiable
and $u\left(Supp(\mathbf{x})\right)$ contains an open ball; and $V$
is twice continuously differentiable with nonzero second-order mixed
partial derivatives. Then it follows that $u:=(u_{a},\;...,u_{K})^{T}$,
with $K=|A|$, is identified over $Supp(\mathbf{x})$ under the following
normalization: $(i)$ (Scale) $\frac{\partial}{\partial x_{b,q}}u_{b}(x_{b},\omega_{j})|_{x_{b}=x_{b}^{*}}\in\{-1,1\}$
for each tuple $(b,q,x_{b}^{*},j)_{j=1}^{J}$ such that $x_{b}^{*}\in Supp(\mathbf{x}_{b})$;
and $(ii)$ (Location) for each $a\in A$,$\vec{u}_{a}(0_{L_{a}})=0$,
where $0_{L_{a}}\in Supp(\mathbf{x}_{a})$.
\end{lem}

\subsection{Identification of mean indirect utility}
\begin{lem}
{[}Theorem 4 in Allen and Rehbeck (2019){]} Let Assumptions 3 and
4 hold. Assume $\vec{u}$ is known, $V$ is everywhere finite, and
let $x,x'\in Supp(\mathbf{x})$. Suppose there is a vector-valued
function $x(t)$ such that, for each $a\in A$, $\omega_{j}\in\Omega$,
$u_{a}\left(x(t),\omega_{j}\right)=tu_{a}(x',\omega_{j})+(1-t)u_{a}(x,\omega_{j})$
and $x(t)\in Supp(\mathbf{x})$ for $t\in[0,1]$. Then it follows
that $V\left(u(x',\omega)\right)-V\left(u(x,\omega)\right)$ and $\bar{D}\left(\mathbb{E}\left[\mathbf{P}|\mathbf{x}=x'\right]\right)-\bar{D}\left(\mathbb{E}\left[\mathbf{P}|\mathbf{x}=x\right]\right)$
are identified. In particular,
\[
V\left(u(x',\omega)\right)-V\left(u(x,\omega)\right)=\intop_{0}^{1}\sum_{j=1}^{J}\sum_{a\in A}\mathbb{E}[\mathbf{P}(a|\omega_{j})|\mathbf{x}=x]\mu(\omega_{j})\left(u_{a}(x_{a}',\omega_{j})-u_{a}(x_{a},\omega_{j})\right)dt.
\]
\end{lem}
\begin{cor}
Let Assumptions 3 and 4 hold. Suppose $\vec{u}$ is known, $V$ is
finite everywhere, and the set $u\left(Supp(\mathbf{x})\right)$ is
convex. It follows that for each $x,x'\in Supp(\mathbf{x})$, $V\left(u(x',\omega)\right)-V\left(u(x,\omega)\right)$
and $\bar{D}\left(\mathbb{E}\left[\mathbf{P}|\mathbf{x}=x'\right]\right)-\bar{D}\left(\mathbb{E}\left[\mathbf{P}|\mathbf{x}=x\right]\right)$
are identified.
\end{cor}
By Lemma 3,

\[
\frac{\partial}{\partial x_{a,p}}V\left(u(x,\omega)\right)=\mathbb{E}[\mathbf{P}(a|\omega_{j})|\mathbf{x}=x]\mu(\omega_{j})\frac{\partial}{\partial x_{a,p}}u_{a}(x_{a},\omega_{j})
\]

By Lemma 6,

\[
\dfrac{\frac{\partial}{\partial x_{a,p}}\mathbb{E}[\mathbf{P}(b|\omega)|\mathbf{x}=x]}{\frac{\partial}{\partial x_{b,q}}\mathbb{E}[\mathbf{P}(a|\omega)|\mathbf{x}=x]}=\dfrac{\frac{\partial}{\partial x_{a,p}}u_{a}(x_{a},\omega_{j})}{\frac{\partial}{\partial x_{b,q}}u_{b}(x_{b},\omega_{j})}.
\]

Then,

\[
\frac{\partial}{\partial x_{a,p}}V\left(u(x,\omega)\right)=\mathbb{E}[\mathbf{P}(a|\omega_{j})|\mathbf{x}=x]\mu(\omega_{j})\dfrac{\frac{\partial}{\partial x_{a,p}}\mathbb{E}[\mathbf{P}(b|\omega_{j})|\mathbf{x}=x]}{\frac{\partial}{\partial x_{b,q}}\mathbb{E}[\mathbf{P}(a|\omega_{j})|\mathbf{x}=x]}\frac{\partial}{\partial x_{b,q}}u_{b}(x_{b},\omega_{j})
\]

Consider $x',x''\in Supp(\mathbf{x})$. Let $x''$ be identical to
$x'$ except for the $(a,p)^{th}$ component, then

\[
\begin{array}{rl}
 & V\left(u(x',\omega)\right)-V\left(u(x,\omega)\right)\\
= & \intop_{x_{a,p}'}^{x_{a,p}''}\mathbb{E}[\mathbf{P}(a|\omega)|\mathbf{x}=x]\mu(\omega_{j})\dfrac{\frac{\partial}{\partial x_{a,p}}\mathbb{E}[\mathbf{P}(b|\omega_{j})|\mathbf{x}=x]}{\frac{\partial}{\partial x_{b,q}}\mathbb{E}[\mathbf{P}(a|\omega_{j})|\mathbf{x}=x]}dx_{a,p}\frac{\partial}{\partial x_{b,q}}u_{b}(x_{b},\omega_{j}).
\end{array}
\]

Recall $\frac{\partial}{\partial x_{b,q}}u_{b}(x_{b},\omega_{j})$
is the scale term at a fixed value $x_{b}$. If $\mathbf{x}_{b}$
is the price of good $b$ and $\frac{\partial}{\partial x_{b,q}}u_{b}(x_{b},\omega_{j})=-1$,
then $V\left(u(x',\omega)\right)-V\left(u(x,\omega)\right)$ can be
interpreted as the change in mean indirect utility in terms of dollars. 

\subsection{Identification of counterfactual bounds}
\begin{lem}
(Counterfactual bounds {[}Theorem 5 in Allen and Rehbeck (2019){]}
Let Assumptions 3 and 4 hold and assume $u$ is known. Let $x^{0}\not\in Supp(\mathbf{x})$
and assume $\mathbb{E}[\mathbf{P}|\mathbf{x}=x^{0}]$ solves

\[
\sup_{P\in\text{Conv}\mathcal{P}}\sum_{j=1}^{J}\sum_{a\in A}P(a|\omega_{j})\mu(\omega_{j})u_{a}(x_{a}^{0},\omega_{j})+\bar{D}(P).
\]
Then, for every integer $S$ and every sequence $x^{1},...,x^{S}\in Supp(\mathbf{x})$,

\[
\begin{array}{rl}
 & \underset{j=1}{\overset{J}{\sum}}\underset{a\in A}{\sum}\mathbb{E}[\mathbf{P}(a|\omega_{j})|\mathbf{x}=x^{0}]\mu(\omega_{j})\left(u_{a}(x_{a}^{0},\omega_{j})-u_{a}(x_{a}^{S-1},\omega_{j})\right)\\
\geq & \underset{j=1}{\overset{J}{\sum}}\underset{a\in A}{\sum}\mu(\omega)\begin{array}[t]{l}
\Bigg[\mathbb{E}[\mathbf{P}(a|\omega_{j})|\mathbf{x}=x^{1}]u_{a}(x^{0},\omega_{j})-\mathbb{E}[\mathbf{P}(a|\omega_{j})|\mathbf{x}=x^{S-1}]u_{a}(x^{S-1},\omega_{j})\\
-\overset{S-2}{\underset{s=1,}{\sum}}\left(\mathbb{E}[\mathbf{P}(a|\omega_{j})|\mathbf{x}=x^{s}]-\mathbb{E}[\mathbf{P}(a|\omega_{j})|\mathbf{x}=x^{s+1}]\right)\mu(\omega_{j})u_{a}(x^{s},\omega_{j})\Bigg].
\end{array}
\end{array}
\]
\end{lem}

\section{Additive separability in latent states}
Define $u_{a}(x_{a},\omega_{j}):=u_{a}(x_{a})+G^{a}(\omega_{j})$,
where $G^{a}:\Omega\mapsto\mathbb{R}$. Treat $\boldsymbol{\omega}$
as a latent variable with known support $\Omega$. Importantly, both
the support of $\boldsymbol{\omega}$ and the functions $\{G^{a}\}_{a\in A}$
are known to the decision maker at the moment of deciding. 

\LyXZeroWidthSpace{}
\begin{assumption}
Suppose the following: $(i)$ $\mathbf{Q}$, $\mathbf{q}$, $\mathbf{x}$,
$\boldsymbol{\omega}$, $\mathbf{D}$, and $\mathbf{T}$ satisfy:

\begin{equation}
\sup_{(Q,q)\in\Lambda}\sum_{\gamma\in Supp(Q)}\sum_{a\in A}Q(\gamma)q(a|\gamma)\left(u_{a}(\mathbf{x}_{a})+\gamma\cdot G^{a}(\boldsymbol{\omega})+\mathbf{E}(a)-\mathbf{T}(\gamma)\right);\label{eq:7}
\end{equation}
$(ii)$ $\mathbf{x}$ and $(\text{\ensuremath{\boldsymbol{\omega}},\ensuremath{\mathbf{E},\mathbf{T})}}$
are independent; $(iii)$ $\mathbf{E}(a)\neq-\infty$ for some $a\in A$
and $\mathbf{T}$ is strictly convex in $\gamma$ and $\mathbf{T}(\gamma)<\infty$
on some $\gamma\in\text{int}\left(\Delta(\Omega)\right)$; $(iv)$
given $G$, the joint probability distribution of $(\text{\ensuremath{\boldsymbol{\omega}},\ensuremath{\mathbf{E},\mathbf{T})}}$
satisfies

\[
\mathbb{P}\left(\left\{ (\text{\ensuremath{\omega},\ensuremath{E,T)}}\in\Omega\times\mathcal{E}\times T\Bigg|\text{rank}\Gamma_{\lambda}(x,\ensuremath{\omega},\ensuremath{E,T)}=\Big|\Gamma(x,\lambda,\omega,\ensuremath{E,T)}\Big|\right\} \right)=1,
\]
for each point $x\in\mathcal{X}$ and each direction $\lambda\in\mathbb{R}^{|\Omega|}$;
and $(v)$ for each $a\in A$, $b\in A\backslash\{a\}$, $G$ satisfies
$G^{a}(\boldsymbol{\omega})\neq G^{b}(\boldsymbol{\omega})$.

\LyXZeroWidthSpace{}
\end{assumption}
\begin{lem}
Let Assumption 5 hold. Then for any fixed $(\omega,E,T)$ in the support,
(\ref{eq:7}) has a unique optimizer $(\hat{Q},\hat{q})$, where at
each $\hat{\gamma}\in Supp(\hat{Q})$, $\hat{q}$ gives a degenerate
distribution over $A$. 
\end{lem}
\LyXZeroWidthSpace{}

Consider the alternative model where $\boldsymbol{\rho}\in\mathcal{P}$
is chosen to satisfy

\begin{equation}
\sup_{\rho\in\mathcal{P}}\sum_{a\in A}\rho(a)\left[u_{a}(\mathbf{x}_{a})+f(\rho,a)\cdot G^{a}(\boldsymbol{\omega})+\mathbf{E}(a)-\mathbf{T}\left(f(\rho,a)\right)\right].\label{eq:8}
\end{equation}

\begin{prop}
(Equivalence) Let Assumption 4 hold. Then, the following propositions
are equivalent:

$(i)$ The posterior-based policy $(\mathbf{Q},\mathbf{q})\in\Lambda$
is the maximizer of (\ref{eq:7}); 

$(ii)$ The SDSC function $\mathbf{P}\in\mathcal{P}$ generated by
$(\mathbf{Q},\mathbf{q})$ is the maximizer of (\ref{eq:8});

$(iii)$ The SDSC function $\boldsymbol{\rho}\in\mathcal{P}$ is the
unique maximizer of (\ref{eq:8}); and

$(iv)$ The pair consisting of the attention and choice distribution
$(\bar{\mathbf{Q}},\bar{\mathbf{q}})$ revealed by $\boldsymbol{\rho}$
is the maximizer of (\ref{eq:7}).
\end{prop}

\subsection{Properties of the model}

Additive separability in states and the independence of covariates
and unobservable heterogeneity ensure the model with latent states
too admits a representative agent. 

\subsubsection{Aggregation}
\begin{lem}
(Theorem 1 in Allen and Rehbeck {[}2019{]}). Let Assumption 5 hold.
Let $x\in Supp(\mathbf{x})$ and suppose $\mathbf{P}$ is $(\mathbf{x},\boldsymbol{\omega},\mathbf{E},\mathbf{T})$-measurable,
and both $\mathbb{E}[\mathbf{P}|\mathbf{x}=x]$ and
\[
\mathbb{E}\left[\sum_{a\in A}\mathbf{P}(a)\left(f(\mathbf{P},a)\cdot G^{a}(\boldsymbol{\omega})+\mathbf{E}(a)-\mathbf{T}\left(f(\mathbf{P},a)\right)\right)\right]
\]
exist and are finite, where the expectations are over the joint distribution
of $(\boldsymbol{\omega},\mathbf{E},\mathbf{T})$. Then it follows
that $(i)$ $\mathbb{E}[\mathbf{P}|\mathbf{x}=x]$ is an optimizer
of the representative agent problem,
\[
\sup_{P\in\mathcal{P}}\sum_{a\in A}P(a)u_{a}(x_{a})+\bar{D}(P),
\]
with

\[
\bar{D}(P)=\sup_{\begin{array}{c}
\{P'\in\mathcal{P}_{(\boldsymbol{\omega},\mathbf{E},\mathbf{T})}|\\
\mathbb{E}[P']=P\}
\end{array}}\mathbb{E}\left[\sum_{a\in A}P'(a)\left(f(P',a)\cdot G^{a}(\boldsymbol{\omega})+\mathbf{E}(a)-\mathbf{T}\left(f(P',a)\right)\right)\right],
\]
where $\mathcal{P}_{(\boldsymbol{\omega},\mathbf{E},\mathbf{T})}$
is the set of $(\boldsymbol{\omega},\mathbf{E},\mathbf{T})$-measurable
functions that map to $\mathcal{P}$. $(ii)$ Define the indirect
utility function of the representative agent problem,

\[
V(v):=\sup_{P\in\mathcal{P}}\sum_{a\in A}P(a)v_{a}+\bar{D}(P).
\]
Thus, $V$ satisfies

\[
V\left(u(x)\right)=\mathbb{E}\left[\sup_{P\in\mathcal{P}}\sum_{a\in A}P(a)\left(u_{a}(x_{a})+f(P,a)\cdot G^{a}(\boldsymbol{\omega})+\mathbf{E}(a)-\mathbf{T}\left(f(P,a)\right)\right)\right].
\]
\end{lem}
\LyXZeroWidthSpace{}

This aggregation property of the model indicates that only conditional
mean stochastic choice data $\mathbb{E}[\mathbf{P}|\mathbf{x}=x]$
is the only data requirement for identification. The fact that the
actual realization of the state needs not to be observed suggests
that market-level data might be used, thus broadening the scope of
the empirical applications of the model.

\subsubsection{Structure}

I enunciate the envelope theorem below and leave the identification
results for the Appendix.
\begin{assumption}
Suppose the following: $(i)$ for each $x\in Supp(\mathbf{x})$, $\mathbb{E}\left[\mathbf{P}|\mathbf{x}=x\right]$
is the unique optimizer of

\[
\sup_{P\in\mathcal{P}}\left\{ \sum_{a\in A}P(a)u_{a}(x_{a})+\bar{D}(P)\right\} ;
\]
and $(ii)$ $\bar{D}:\mathcal{P}\mapsto\mathbb{R}\cup\{-\infty\}$
is concave, upper semi-continuous, and finite at some $P\in\mathcal{P}$.
\end{assumption}
\begin{lem}
Let Assumption 6 hold. Then, 

\[
\mathbb{E}[\mathbf{P}|\mathbf{x}=x]=\nabla V\left(u(x)\right).
\]
\end{lem}
\begin{lem}
Let Assumption 6 hold and let $V$ be twice continuously differentiable
in a neighbourhood of $v\in\mathbb{R}^{|A|}$. Then, the following
conditions are satisfied: $(i)$ $\lambda^{T}\nabla^{2}V(v)\lambda\geq0$,
for all $\lambda\in\mathbb{R}^{|A|}$; and $(ii)$ $\frac{\partial^{2}}{\partial v_{a}\partial v_{b}}V(v)=\frac{\partial^{2}}{\partial v_{b}\partial v_{a}}V(v)$,
for all $a,b\in A$.
\end{lem}
\begin{lem}
Let Assumption 6 hold. Define

\[
\bar{D}_{\mathcal{P}}(P):=\begin{cases}
\bar{D}(P), & \text{if }P\in\mathcal{P}\\
-\infty, & \text{otherwise.}
\end{cases}
\]
If $\bar{D}_{\mathcal{P}}$ is differentiable at $\mathbb{E}[\mathbf{P}|\mathbf{x}=x]$,
then for each $a\in A$,

\[
u_{a}(x_{a})=-\frac{\partial}{\partial P_{a}}\bar{D}\left(\mathbb{E}[\mathbf{P}|\mathbf{x}=x]\right).
\]
\end{lem}

\section{Concluding remarks}
I present a theoretical model with additively separable latent heterogeneity
that describes the behavior of a population of rationally inattentive
decision makers. Under some regularity conditions, I show that this
model is observationally equivalent to a state-dependent stochastic
choice model constrained by attention costs. In the interest of learning
how demand responds to changes in attributes, I include covariates
as arguments in the utility indices. Assuming regressors and unobservable
heterogeneity are independent, I show that my model admits a representative
agent. This aggregation property, together with the structure of the
model, allows me to to identify structural and counterfactual parameters
when the conditional mean SDSC function is observable, that is, when
the econometrician observes conditional probabilities given states
and covariates. In particular, I identify how attributes shift the
desirability of different goods, (a measure of) welfare, factual changes
in welfare, and bounds on counterfactual probabilities of choice.
Further assuming utility indices are additively separable in (latent)
states, I establish identification with conditional mean stochastic
choice data. In the latter case, as the analyst does not need to observe
the realization of the state, the model can be used for empirical
applications using market-level data.

I close by mentioning extensions to the paper that I will pursue in
further research:

$(i)$ Some applications require the states to be continuous. I will
generalize the aggregation theorem so as to accomodate for infinite
states.

$(ii)$ Although utility indices are identified up to location and
scale and unobservable heterogeneity is recovered, the components
of the latter, that is, latent preferences and latent attention costs
are not separately identified. I will place bounds on both components
of latent heterogeneity by deriving the empirical content of the model,
under the assumption that the econometrician observes exogenous variation
in choice sets. This strategy might also serve for placing bounds
on choice frequencies after a counterfactual change in choice sets.

$(iii)$ So far, I assumed a known, homogeneous prior. I will let
the prior be a latent random variable. Using again the empirical content
of the model, I will build a finite mixture model based on equivalence
classes, i.e. partitions of the space of subjective priors that are
equivalent in terms of revealed preference relations.

$(iv)$ I will generalize my model to include other types of costly
information acquisition, such as wishful thinking.

\newpage{}

\section*{References}

Aguiar, V. H., Boccardi, M. J., Kashaev, N., \& Kim, J. (2023). Random
utility and limited consideration. Quantitative Economics, 14(1),
71-116.

Allen, R., \& Rehbeck, J. (2019). Identification with additively separable
heterogeneity. Econometrica, 87(3), 1021-1054.

Bénabou, R., \& Tirole, J. (2016). Mindful economics: The production,
consumption, and value of beliefs. Journal of Economic Perspectives,
30(3), 141-164.

Brown, Z. Y., \& Jeon, J. (2020). Endogenous information and simplifying
insurance choice. Mimeo, University of Michigan.

Caplin, A., \& Dean, M. (2015). Revealed preference, rational inattention,
and costly information acquisition. American Economic Review, 105(7),
2183-2203.

Caplin, A., Dean, M., \& Leahy, J. (2019). Rational inattention, optimal
consideration sets, and stochastic choice. The Review of Economic
Studies, 86(3), 1061-1094.

Caplin, A., Dean, M., \& Leahy, J. (2022). Rationally inattentive
behavior: Characterizing and generalizing Shannon entropy. Journal
of Political Economy, 130(6), 1676-1715.

Caplin, A., \& Leahy, J. V. (2019). Wishful thinking (No. w25707).
National Bureau of Economic Research.

Fosgerau, M., Melo, E., De Palma, A., \& Shum, M. (2020). Discrete
choice and rational inattention: A general equivalence result. International
economic review, 61(4), 1569-1589.

Joo, J. (2023). Rational inattention as an empirical framework for
discrete choice and consumer-welfare evaluation. Journal of Marketing
Research, 60(2), 278-298.

Kovach, M. (2020). Twisting the truth: Foundations of wishful thinking.
Theoretical Economics, 15(3), 989-1022.

Liao, M. (2024). Identification of a rational inattention discrete
choice model. Journal of Econometrics, 240(1), 105670.

Ma\'{c}kowiak, B., Mat\v{e}jka, F., \& Wiederholt, M. (2023). Rational
inattention: A review. Journal of Economic Literature, 61(1), 226-273.

Mat\v{e}jka, F., \& McKay, A. (2015). Rational inattention to discrete
choices: A new foundation for the multinomial logit model. American
Economic Review, 105(1), 272-298.

Rockafellar, R. T. (1970). Convex Analysis. Princeton Math. Series,
28.\newpage{}

\section*{Appendix A --Main proofs}

\subsection*{Proof of Proposition 1}
Before proving this proposition, I introduce some necessary definitions
and intermediate results. Fix any $(E,T).$ Define: the concave conjugate
of $N$ in the direction $\theta$,
\[
N^{*}(\theta):=\sup_{\gamma\in\Delta(\Omega)}N(\gamma)-\theta\cdot\gamma;
\]
the lower concave envelope of $N$,

\[
\ensuremath{\underline{N}}:=\text{inf}\left\{ \tilde{N}:\Delta(\Omega)\mapsto\mathbb{R}|\tilde{N}\text{-concave},\;\tilde{N}\geq N\right\} ;
\]
and the set of directions $\theta$ at which the prior lies in the
convex hull of the set of supported posteriors, 
\[
\varTheta:=\left\{ \theta\in\mathbb{R}^{|\Omega|-1}\Big|\mu\in\text{Conv}\varGamma(\theta)\right\} .
\]

\begin{lem}
(Lagrangian Lemma {[}Lemma 1 in Caplin, Dean and Leahy (2022){]}.
Fix any $(D,T)$. $(\hat{Q},\hat{q})\in\Lambda$ is an optimizer of
(\ref{eq:1}) iff there exists $\theta\in\mathbb{R}^{|\Omega|-1}$
such that, for each $\hat{\gamma}\in Supp(\hat{Q})$ and $a\in A$
with $q(a|\hat{\gamma})>0$, the following inequality holds

\[
N^{a}(\hat{\gamma})-\theta\cdot\hat{\gamma}\geq N^{b}(\gamma')-\theta\cdot\gamma',\;\text{for all }b\in A,\;\gamma'\in\Delta(\Omega).
\]
\end{lem}
\begin{lem}
(Carathéodory's Theorem {[}Theorem 17.1 in Rockafellar (1970){]})
Let $S$ be any set of points and directions in $\mathbb{R}^{|\Omega|-1}$.
Then $\mu\in\text{Conv}(S)$ iff $\mu$ can be expressed as a convex
combination of at most $|\Omega|$ of the points and directions in
$S$.
\end{lem}
\begin{cor}
{[}Corollary 17.1.5 in Rockafellar (1970){]} Let $N$ be a function
from $\mathbb{R}^{|\Omega|-1}$ to $(-\infty,+\infty]$. Then,

\[
\text{Conv}N(\mu)=\inf\left\{ \sum_{n=1}^{N}Q^{(n)}N(\gamma^{(n)})\Bigg|\sum_{n=1}^{N}Q^{(n)}\gamma^{(n)}=\mu\right\} ,
\]
where $\text{Conv}N(\mu)$ denotes the convex hull of the function
$N$at $\mu$.
\end{cor}
\begin{lem}
Fix any direction $\theta\in\mathbb{R}^{|\Omega|-1}$. By definition
of $N^{*}$, $\varGamma$, and $\delta^{*}$, the following statements
are equivalent:

$(i)$ $\gamma\in\Delta(\Omega)$ satisfies $N(\gamma)-\theta\cdot\gamma\geq N(\gamma')-\theta\cdot\gamma'$,
for all $\gamma'\in\Delta(\Omega)$;

$(ii)$ $\gamma\in\Delta(\Omega)$ satisfies $N(\gamma)-\theta\cdot\gamma=N^{*}(\theta)$;

$(iii)$ $\gamma\in\varGamma(\theta)$; and

$(iv)$ $\left(\gamma,N(\gamma)\right)\in\Delta(\Omega)\times\mathbb{R}$
satisfies $N(\gamma)-\theta\cdot\gamma=\delta^{*}\left((1,-\theta)|\text{hyp}N\right)$.
\end{lem}
\begin{proof}
First, I prove uniqueness of optimal attention. Consider the lower
concave envelope of $N$ at the prior,
\[
\begin{array}[t]{rcl}
\underline{N}(\mu) & = & \sup_{\{Q\in\Delta\left(\Delta(\Omega)\right):\sum_{\gamma\in Supp(Q)}Q(\gamma)\gamma=\mu\}}\sum_{\gamma\in Supp(Q)}Q(\gamma)N(\gamma)\\
 & = & \sup_{Q\in\Delta\left(\varGamma(\theta)\right),\theta\in\varTheta}\sum_{\gamma\in Supp(Q)}Q(\gamma)N(\gamma)\\
 & = & \sup_{\gamma\in\varGamma(\theta),\theta\in\varTheta}\delta^{*}\left((1,-\theta)|\text{hyp}N\right)+\theta\cdot\gamma.
\end{array}
\]

The first equality, that is, the lower concave envelope of $N$ at
$\mu$ being equal to the optimization problem in (\ref{eq:3}), is
implied by Corollary 4. From Lemma 15 (Lagrangian lemma), it follows
that $\hat{Q}\in\mathcal{Q}$ is a maximizer only provided that, for
some $\theta\in\mathbb{R}^{|\Omega|-1}$, each $\hat{\gamma}\in Supp(\hat{Q})$
satisfies
\[
N(\hat{\gamma})-\theta\cdot\hat{\gamma}\geq N(\gamma')-\theta\cdot\gamma',\;\text{for all }\gamma'\in\Delta(\Omega).
\]
Therefore, by equivalence between $(i)$, $(ii)$, and $(iii)$ in
Lemma 16, $\hat{Q}$ maximizing (\ref{eq:3}) is constrained to lie
in the probability simplex over $\varGamma(\theta)$ for some $\theta\in\mathbb{R}^{|\Omega|-1}$.
Moreover, in order for $\hat{Q}$ to be Bayes-consistent, it is necessarily
the case that $\theta$ lives in $\varTheta$, which implies the equality
in the second line. An optimal attention $\hat{Q}$ is such that,
by definition of $N^{*}(\theta)$, each optimal posterior $\hat{\gamma}\in Supp(\hat{Q})\subseteq\varGamma(\theta)$
satisfies $N^{*}(\theta)+\theta\cdot\hat{\gamma}=N(\hat{\gamma})$
for some $\theta\in\varTheta$. The equality in the third line follows
from the equivalence between $(iii)$ and $(iv)$ in Lemma 16.

Note that, due to the lower concave envelope of $N$ at $\mu$ being
a convex combination of all values $N(\hat{\gamma})$ where each $\hat{\gamma}$
is an optimal posterior, it follows that it is linear in $\gamma$,
thus being differentiable. As a result, by equality between $\underline{N}(\mu)$
and the supporting hyperplane in the direction $(1,-\theta)$, there
is a unique optimal direction $\hat{\theta}=\partial\underline{N}(\mu)$.
By Lemma 16 (Carathéodory's Theorem), $\mu\in\text{Conv}\varGamma\left(\partial\underline{N}(\mu)\right)$
entails it can be written as a convex combination of at most $|\Omega|$
posteriors in $\varGamma\left(\partial\underline{N}(\mu)\right)$.
Since Assumption $1$ $(iii)$ (full column rank) implies that $\varGamma\left(\partial\underline{N}(\mu)\right)$
consists of $\Big|\varGamma\left(\partial\underline{N}(\mu)\right)\Big|\leq|\Omega|$
linearly independent posteriors, such convex combination is unique
and equal to a fully-supported distribution $\hat{Q}\in\Delta\left(\varGamma\left(\partial\underline{N}(\mu)\right)\right)$,
the optimal attention.

\LyXZeroWidthSpace{}

In what follows, I show that optimal choice is unique and gives a
degenerate distribution at each optimal posterior. Define the set
of alternatives that yield the maximum net expected utility according
to a given posterior belief $\gamma$,
\[
A(\gamma):=\arg\max_{a\in A}N^{a}(\gamma).
\]

An optimal choice policy $\hat{q}$ is such that, for each posterior
supported by the optimal attention, $\hat{\gamma}\in Supp\hat{Q}$,
it is chosen so that
\[
\hat{q}(\hat{\gamma})\in\arg\max_{q\in\Delta(A)}\sum_{a\in A}q(a|\hat{\gamma})N^{a}(\hat{\gamma}).
\]

By Assumption 1 $(iv)$, for any $a,b\in A$, with $a\neq b$, $u_{a}\neq u_{b}$,
which implies $\partial N^{a}(\gamma)\neq\partial N^{b}(\gamma)$
at each $\gamma\in\Delta(\Omega)$. As a result, for any $\gamma\in\Delta(\Omega)$,
$|A(\gamma)|=1$. For each $\hat{\gamma}\in Supp(\hat{Q})$, $\hat{q}(\gamma)$
is therefore a degenerate distribution that assigns probability mass
1 to a unique $a\in A$ such that $\{a\}=A(\hat{\gamma})$. Due to
its degenerateness, the optimal choice is unique.
\end{proof}

\subsection*{Proof of Proposition 2}
\begin{proof}
Fix any $(E,T)$ in the support and consider the following optimization
problem

\[
\begin{array}{rl}
 & \underline{N}(\mu)\\
= & \underset{\rho\in\mathcal{P}}{\sup}\underset{a\in A}{\sum}\rho(a)N\left(f(\rho;a)\right)\\
= & \underset{\rho\in\mathcal{Q}}{\sup}\underset{\{a\in A|\exists f(\rho;a)\in Supp(\rho)\}}{\sum}Q\left(f(\rho;a)\right)N^{a}\left(f(\rho;a)\right).
\end{array}
\]

The expressions in the first and second line being equivalent means
the lower concave envelope of $N$ at $\mu$ is attained by the optimization
problem in (\ref{eq:4}). This follows from Corollary 4 and the fact
that $\rho$ averages the values $\left(f(\rho;a)\right)_{a\in A}$
to the prior.\footnote{$\begin{array}[t]{rcl}
\sum_{a\in A}\rho(a)f\left(\rho;a\right) & = & \sum_{a\in A}\rho(a)\left(f_{\omega_{1}}\left(\rho;a\right),\;....,\;f_{\omega_{J}}\left(\rho;a\right)\right)^{T}\\
 & = & \left(\sum_{a\in A}\rho(a)f_{\omega_{1}}\left(\rho;a\right),\;....,\;\sum_{a\in A}\rho(a)f_{\omega_{J}}\left(\rho;a\right)\right)^{T}\\
 & = & \left(\sum_{a\in A}\rho(a)\frac{\rho(a|\omega_{1})\mu(\omega_{1})}{\rho(a)},\;....,\;\sum_{a\in A}\rho(a)\frac{\rho(a|\omega_{J})\mu(\omega_{J})}{\rho(a)}\right)^{T}\\
 & = & \left(\mu(\omega_{1})\sum_{a\in A}\rho(a|\omega_{1}),\;....,\;\mu(\omega_{J})\sum_{a\in A}\rho(a|\omega_{J})\right)^{T}\\
 & = & \left(\mu(\omega_{1}),\;....,\;\mu(\omega_{J})\right)^{T}\\
 & = & \mu.
\end{array}$} The equalitiy in the third line follows directly from Corollary 1.
By Lemma 16 (Carathéodory's Theorem), $\mu$ being in the convex hull
of the set $\left(f(\rho;a)\right)_{a\in A}$, entails $\mu$ can
be written as a convex combination of at most $|\Omega|-1$ values
in the set, restricting the cardinality of this set to $\Big|\left(f(\rho;a)\right)_{a\in A}\Big|\leq|\Omega|-1$.
The full column rank condition restricts the cardinality of the set
of vectors $\left(f(\hat{\rho};a)\right)_{a\in Supp(\hat{\rho})}$,
and hence, the cardinality of the support of $\hat{\rho}$, to be
at most $|\Omega$| for optimal $\hat{\rho}$. Therefore, the convex
combination satisfying $\sum_{a\in Supp(\rho)}\hat{\rho}(a)f(\hat{\rho};a)=\mu$
is unique. In consequence, the optimal SDSC function $\hat{\rho}$
is unique.
\end{proof}

\subsection*{Proof of Theorem 1}
\begin{proof}
Fix any $(E,T)$ in the support and consider the following sequence
of equalities,

\[
\begin{array}[t]{rl}
 & \underline{N}(\mu)\\
= & \sup_{Q\in\mathcal{Q}}\sum_{\{a\in A|\exists\gamma\in Supp(Q):\gamma=\gamma^{a}\}}Q(\gamma^{a})N^{a}(\gamma^{a})\\
= & \sup_{P\in\mathcal{P}}\sum_{a\in A}P(a)N^{a}(\gamma^{a})\\
= & \sup_{\rho\in P}\sum_{a\in A}\rho(a)N^{a}\left(f(\rho,a)\right)\\
= & \sup_{\rho\in P}\sum_{a\in A}\rho(a)N^{a}\left(\bar{\gamma}^{a}\right)\\
= & \sup_{\bar{Q}\in\mathcal{Q}}\sum_{\{a\in A|\exists\gamma\in Supp(\bar{Q}):\gamma=\bar{\gamma}^{a}\}}\bar{Q}\left(\bar{\gamma}^{a}\right)N^{a}\left(\bar{\gamma}^{a}\right)\\
= & \underline{N}(\mu).
\end{array}
\]

It is known from proof of Proposition 1 that the lower concave envelope
of $N$ at the prior equals the maximized value function in (\ref{eq:1}).
The equality in the second line follows from (\ref{eq:1}) being equivalent
to (\ref{eq:3}). The equality in the third line is implied by the
definition of generated SDSC function $P\in\mathcal{P}$. $\gamma^{a}\in Supp(Q)$
satisfying $\gamma^{a}=f(P;a)$ entails the equality in the fourth
line. The posterior-based policy $(\hat{Q},\hat{q})\in\Lambda$ that
optimizes (\ref{eq:1}) generates a SDSC function $\hat{P}$ that
optimizes (\ref{eq:4}).

The equality in the fifth line follows from the definition of revealed
posteriors. The equality in the sixth line follows from the definition
of revealed attention and the fact that it averages revealed posteriors
to the prior.\footnote{$\sum_{\{a\in A|\gamma\in Supp(\bar{Q}):\gamma=\bar{\gamma}^{a}\}}\bar{Q}(\bar{\gamma}^{a})\bar{\gamma}^{a}=\sum_{a\in Supp(\rho)}\rho(a)\left(\frac{\rho(a|\omega)\mu(\omega_{1})}{\rho(a)},\;...,\;\frac{\rho(a|\omega)\mu(\omega_{J})}{\rho(a)}\right)=\mu.$}
By Corollary 4, the optimization problem over Bayes-consistent revealed
attention policies attains the lower concave envelope of $N$ at $\mu$.
The SDSC function $\hat{\rho}\in\mathcal{P}$ that optimizes (\ref{eq:4})
reveals a posterior based policy $(\bar{Q},\bar{q})$ that optimizes
(\ref{eq:1}).

\newpage{}
\end{proof}

\section*{Appendix B --Proofs in Allen and Rehbeck (2019)}

\subsection*{Proof of Lemma 2 (Theorem 1 in Allen and Rehbeck, 2019)}
\begin{proof}
Below I follow the proof of Theorem 1 in Allen and Rehbeck (2019).

Consider the following cycle of inequalities beginning with the expected
value of the pointwise optimization problem,\LyXZeroWidthSpace{}

\[
\begin{array}{rl}
 & \mathbb{E}\left[\underset{P\in\mathcal{P}}{\sup}\underset{a\in A}{\sum}P(a)\left(f(P,a)\cdot u(x_{a})+\mathbf{E}(a)-\mathbf{T}\left(f(P,a)\right)\right)\right]\\
= & \underset{j=1}{\overset{J}{\sum}}\underset{a\in A}{\sum}\mathbb{E}[\mathbf{P}(a|\omega_{j})|\mathbf{x}=x]\mu(\omega_{j})u_{a}(x_{a},\omega_{j})+\mathbb{E}\left[\underset{a\in A}{\sum}\mathbf{P}(a)\left(\mathbf{E}(a)-\mathbf{T}\left(f(\mathbf{P},a)\right)\right)\right]\\
\leq & \begin{array}[t]{l}
\underset{j=1}{\overset{J}{\sum}}\underset{a\in A}{\sum}\mathbb{E}[\mathbf{P}(a|\omega_{j})|\mathbf{x}=x]\mu(\omega_{j})u_{a}(x_{a},\omega_{j})\\
+\underset{\left\{ P'\in\mathcal{P}_{(\mathbf{E},\mathbf{T})}|\mathbb{E}[P']=\mathbb{E}[\mathbf{P}|\mathbf{x}=x]\right\} }{\sup}\mathbb{E}\left[\underset{a\in A}{\sum}P'(a)\left(\mathbf{E}(a)-\mathbf{T}\left(f(P',a)\right)\right)\right]\Bigg\}
\end{array}\\
\leq & \begin{array}[t]{ll}
\underset{P\in\text{Conv}(\mathcal{P})}{\sup} & \underset{j=1}{\overset{J}{\sum}}\underset{a\in A}{\sum}P(a|\omega_{j})\mu(\omega_{j})u_{a}(x_{a},\omega_{j})\\
 & +\underset{\left\{ P'\in\mathcal{P}_{(\mathbf{E},\mathbf{T})}\Big|\mathbb{E}[P']=P\right\} }{\sup}\begin{array}[t]{l}
\Bigg\{\mathbb{E}\left[\underset{a\in A}{\sum}P'(a)\left(\mathbf{E}(a)-\mathbf{T}\left(f(P',a)\right)\right)\right]\Bigg\}\end{array}
\end{array}\\
= & \underset{P\in\text{Conv}(\mathcal{P})}{\sup}\underset{\begin{array}{c}
\{P'\in\mathcal{P}_{(\mathbf{E},\mathbf{T})}|\\
\mathbb{E}[P']=p\}
\end{array}}{\sup}\mathbb{E}\left[\underset{a\in A}{\sum}P(a)\left(f(P,a)\cdot u_{a}(x_{a})+\mathbf{E}(a)-\mathbf{T}\left(f(P',a)\right)\right)\right]\\
\leq & \underset{P'\in\mathcal{P}_{(\mathbf{E},\mathbf{T})}}{\sup}\mathbb{E}\left[\underset{a\in A}{\sum}P'(a)\left(f(P',a)\cdot u_{a}(x_{a})+\mathbf{E}(a)-\mathbf{T}\left(f(P',a)\right)\right)\right]\\
\leq & \mathbb{E}\left[\underset{P\in\mathcal{P}}{\sup}\underset{a\in A}{\sum}P(a)\left(f(P,a)\cdot u_{a}(x_{a})+\mathbf{E}(a)-\mathbf{T}\left(f(P,a)\right)\right)\right].
\end{array}
\]

The equality in the second line obtains by $\mathbf{P}$ being the
maximizer of $(2)$ at $(x,\mathbf{E},\mathbf{T})$ and distributing
the expectation. The inequality in the third line follows from the
assumptions that $\mathbf{P}$ at $x$ is $(\mathbf{E},\mathbf{T})$-measurable
and $\mathbb{E}[\mathbf{P}|\mathbf{x}=x]$ exists. Since $\mathbf{P}\in\mathcal{P}$
at each $(E,T)$ in the support and it is assumed that $\mathbb{E}[\mathbf{P}|\mathbf{x}=x]$
is finite, then $\mathbb{E}[\mathbf{P}|\mathbf{x}=x]\in\mathcal{P}$,
thus explaining the inequality in the fourth line. The equality in
the fifth line results from $\sum_{j=1}^{J}\sum_{a\in A}P(a|\omega_{j})\mu(\omega_{j})u_{a}(x_{a},\omega_{j})$
being a constant term relative to $P'$ and $(\mathbf{E},\mathbf{T})$.
The inequality in the sixth line is implied by relaxing constrains
in the feasibility set in the optimization problem in the previous
line. At each $(\mathbf{E},\mathbf{T})$ in the support, the optimizer
in the pointwise problem in the last line satisfies

\[
\begin{array}{rl}
 & \underset{\omega\in\Omega}{\sum}\underset{a\in A}{\sum}\mathbf{P}(a|\omega)\mu(\omega)u_{a,\omega}(x_{a})+\underset{a\in A}{\sum}\mathbf{P}(a)\left(\mathbf{E}(a)-\mathbf{T}\left(f(\mathbf{P},a)\right)\right)\\
\geq & \underset{\omega\in\Omega}{\sum}\underset{a\in A}{\sum}P'(a|\omega)\mu(\omega)u_{a,\omega}(x_{a})+\underset{a\in A}{\sum}P'(a)\left(\mathbf{E}(a)-\mathbf{T}\left(f(P',a)\right)\right),
\end{array}
\]
for each $P\in\mathcal{P}_{(\mathbf{E},\mathbf{T})}$, which implies
the inequality in the seventh line. Because the chain of inequalities
is such that the expression in the first and last line are identical,
it follows that all the inqualities hold with equality. In particular,
note that by equality between the third and fourth lines, the mean
optimizer of the pointwise problem $\mathbb{E}[\mathbf{P}|\mathbf{x}=x]$
solves the representative agent problem,

\[
\begin{array}[t]{rl}
 & \underset{j=1}{\overset{J}{\sum}}\underset{a\in A}{\sum}\mathbb{E}[\mathbf{P}(a|\omega_{j})|\mathbf{x}=x]\mu(\omega_{j})u_{a}(x_{a},\omega_{j})+\bar{D}(\mathbb{E}[\mathbf{P}|\mathbf{x}=x])\\
= & \underset{P\in\text{Conv}(\mathcal{P})}{\sup}\underset{j=1}{\overset{J}{\sum}}\underset{a\in A}{\sum}P(a|\omega_{j})\mu(\omega_{j})u_{a}(x_{a},\omega_{j})+\bar{D}(P).
\end{array}
\]
Moreover, by equality between the fourth and seventh lines, the value
function in the representative agent problem equals the mean maximized
value in the pointwise problem,
\[
\begin{array}{rl}
 & \underset{P\in\text{Conv}(\mathcal{P})}{\sup}\underset{j=1}{\overset{J}{\sum}}\underset{a\in A}{\sum}P(a|\omega_{j})\mu(\omega_{j})u_{a}(x_{a},\omega_{j})+\bar{D}(P)\\
= & \mathbb{E}\left[\underset{P\in\mathcal{P}}{\sup}\underset{j=1}{\overset{J}{\sum}}\underset{a\in A}{\sum}P(a|\omega_{j})\mu(\omega_{j})u_{a}(x_{a},\omega_{j})+\underset{a\in A}{\sum}P(a)\left(\mathbf{E}(a)-\mathbf{T}\left(f(P,a)\right)\right)\right].
\end{array}
\]

\end{proof}

\subsection*{Proof of Lemma 3}
\begin{proof}
Consider the following sequence of equalities,

\[
\begin{array}[t]{rl}
 & \underset{j=1}{\overset{J}{\sum}}\underset{a\in A}{\sum}\mathbb{E}\left[\mathbf{P}(a|\omega_{j})|\mathbf{x}=x\right]\mu(\omega_{j})u_{a}(x_{a},\omega_{j})+\bar{D}(\mathbb{E}\left[\mathbf{P}|\mathbf{x}=x\right])\\
= & \underset{P\in\text{Conv}(\mathcal{P})}{\sup}\underset{j=1}{\overset{J}{\sum}}\underset{a\in A}{\sum}P(a|\omega_{j})\mu(\omega_{j})u_{a}(x_{a},\omega_{j})+\bar{D}\left(P\right)\\
= & V\left(\vec{u}(x)\right)\\
= & \mathbb{E}\left[\underset{P\in\mathcal{P}}{\sup}\underset{j=1}{\overset{J}{\sum}}\underset{a\in A}{\sum}P(a|\omega_{j})\mu(\omega_{j})u_{a}(x_{a},\omega_{j})+\underset{a\in A}{\sum}P(a)\left(\mathbf{E}(a)-\mathbf{T}\left(f(P,a)\right)\right)\right]\\
= & \underset{j=1}{\overset{J}{\sum}}\underset{a\in A}{\sum}\mathbb{E}\left[\mathbf{P}(a|\omega_{j})|\mathbf{x}=x\right]\mu(\omega_{j})u_{a}(x_{a},\omega_{j})+\mathbb{E}\left[\underset{a\in A}{\sum}\mathbf{P}(a)\left(\mathbf{E}(a)-\mathbf{T}\left(f(\mathbf{P},a)\right)\right)\right],
\end{array}
\]
where the first equivalence follows from $\mathbb{E}\left[\mathbf{P}|\mathbf{x}=x\right]$
being the maximizer of the problem of the representative agent and
the equivalence in the third line is implied by the definition of
the average indirect utility. Lemma 2 $(ii)$ entails the equality
in the fourth line. The equality in the fifth line follows from distributing
the expectation.

Hence, taking the partial derivative of $V$ at $\vec{u}(x)$ with
respect to its $(a,j)^{th}$ component,

\[
\frac{\partial}{\partial v_{a,j}}V\left(u(x,\omega)\right)\Bigg|_{v=u(x,\omega)}=\mathbb{E}\left[\mathbf{P}(a|\omega_{j})|\mathbf{x}=x\right]\mu(\omega_{j}).
\]
\end{proof}

\subsection*{Proof of Lemma 5}
\begin{proof}
Recall $\mathbb{E}[\mathbf{P}|\mathbf{x}=x]$ optimizes the representative
agent problem at $x$, thus attaining the average indirect utility
at $u(x)$. Differentiate $V\left(u(x,\omega)\right)$ with respect
to any $P(a|\omega_{j})$,

\[
\begin{array}{rcl}
\frac{\partial}{\partial P(a|\omega_{j})}V\left(u(x,\omega)\right) & = & \mu(\omega_{j})u_{a}(x_{a},\omega_{j})+\frac{\partial}{\partial P(a|\omega_{j})}\bar{D}_{\text{Conv\ensuremath{\mathcal{P}}}}(P)\Bigg|_{P=\mathbb{E}[\mathbf{P}|\mathbf{x}=x]}\\
 & = & 0.
\end{array}
\]
where the second equality comes from differentiating the value function
evaluated at the optimizer. Hence,

\[
\mu(\omega_{j})u_{a}(x_{a},\omega_{j})=-\frac{\partial}{\partial P(a|\omega_{j})}\bar{D}_{\text{Conv\ensuremath{\mathcal{P}}}}(P)\Bigg|_{P=\mathbb{E}[\mathbf{P}|\mathbf{x}=x]}.
\]
\end{proof}

\subsection*{Proof of Lemma 6}
\begin{proof}
By Lemma 4 $(ii)$ (symmetry), for any $\omega_{j}\in\Omega$, $a,b\in A$, 

\[
\frac{\partial^{2}}{\partial u_{a}(x_{a},\omega_{j})\partial u_{b}(x_{b},\omega_{j})}V\left(u(x,\omega)\right)=\frac{\partial^{2}}{\partial u_{b}(x_{b},\omega_{j})\partial u_{a}(x_{a},\omega_{j})}V\left(u(x,\omega)\right).
\]
By Lemma 3 (Roy's identity),

\[
\mathbb{E}[\mathbf{P}(a|\omega_{j})|\mathbf{x}=x]\mu(\omega_{j})=\frac{\partial}{\partial u_{a}(x_{a},\omega_{j})}V\left(u(x,\omega)\right)\;\text{and}
\]
\[
\mathbb{E}[\mathbf{P}(b|\omega_{j})|\mathbf{x}=x]\mu(\omega_{j})\frac{\partial}{\partial u_{b}(x_{b},\omega)}V\left(u(x,\omega)\right).
\]
 Hence,

\[
\begin{array}{rcl}
\frac{\partial}{\partial u_{b}(x_{b},\omega_{j})}\mathbb{E}[\mathbf{P}(a|\omega_{j})|\mathbf{x}=x]\mu(\omega_{j}) & = & \frac{\partial^{2}}{\partial u_{b}(x_{b},\omega_{j})\partial u_{a}(x_{a},\omega_{j})}V\left(u(x,\omega)\right)\\
 & = & \frac{\partial^{2}}{\partial u_{a}(x_{a},\omega)\partial u_{b}(x_{b},\omega_{j})}V\left(u(x,\omega)\right)\\
 & = & \frac{\partial}{\partial u_{a}(x_{a},\omega_{j})}\mathbb{E}[\mathbf{P}(b|\omega_{j})|\mathbf{x}=x]\mu(\omega_{j}).
\end{array}
\]
Then, 

\[
\frac{\partial}{\partial u_{b}(x_{b},\omega_{j})}\mathbb{E}[\mathbf{P}(a|\omega_{j})|\mathbf{x}=x]=\frac{\partial}{\partial u_{a}(x_{a},\omega_{j})}\mathbb{E}[\mathbf{P}(b|\omega_{j})|\mathbf{x}=x].
\]
Taking partial derivatives

\[
\frac{\partial}{\partial x_{b,q}}\mathbb{E}[\mathbf{P}(a|\omega_{j})|\mathbf{x}=x]=\frac{\partial}{\partial u_{b}(x_{b},\omega_{j})}\mathbb{E}[\mathbf{P}(a|\omega_{j})|\mathbf{x}=x]\frac{\partial}{\partial x_{b,q}}u_{b}(x_{b},\omega_{j}),
\]

\[
\frac{\partial}{\partial x_{a,p}}\mathbb{E}[\mathbf{P}(b|\omega_{j})|\mathbf{x}=x]=\frac{\partial}{\partial u_{a}(x_{a},\omega_{j})}\mathbb{E}[\mathbf{P}(b|\omega_{j})|\mathbf{x}=x]\frac{\partial}{\partial x_{a,p}}u_{a}(x_{a},\omega_{j}).
\]
Thus, by $\frac{\partial}{\partial u_{b}(x_{b},\omega_{j})}\mathbb{E}[\mathbf{P}(a|\omega_{j})|\mathbf{x}=x]=\frac{\partial}{\partial u_{a}(x_{a},\omega_{j})}\mathbb{E}[\mathbf{P}(b|\omega_{j})|\mathbf{x}=x]$,

\[
\dfrac{\frac{\partial}{\partial x_{a,p}}\mathbb{E}[\mathbf{P}(b|\omega_{j})|\mathbf{x}=x]\Big|_{x=x^{*}}}{\frac{\partial}{\partial x_{b,q}}\mathbb{E}[\mathbf{P}(a|\omega_{j})|\mathbf{x}=x]\Big|_{x=x^{*}}}=\dfrac{\frac{\partial}{\partial x_{a,p}}u_{a}(x_{a},\omega_{j})\Big|_{x_{a}=x_{a}^{*}}}{\frac{\partial}{\partial x_{b,q}}u_{b}(x_{b},\omega_{j})\Big|_{x_{b}=x_{b}^{*}}}.
\]
\end{proof}

\subsection*{Proof of Lemma 12}
\begin{proof}
Consider the following sequence of equivalent expressions,

\[
\begin{array}[t]{rl}
 & \underset{a\in A}{\sum}\mathbb{E}[\mathbf{P}(a)|\mathbf{x}=x]u_{a}(x_{a})+\bar{D}\left(\mathbb{E}[\mathbf{P}|\mathbf{x}=x]\right)\\
= & \underset{P\in\text{Conv}(\mathcal{P})}{\sup}\underset{a\in A}{\sum}P(a)u_{a}(x_{a})+\bar{D}\left(P\right)\\
= & V\left(u(x)\right)\\
= & \mathbb{E}\left[\underset{P\in\mathcal{P}}{\sup}\underset{a\in A}{\sum}P(a)\left(u_{a}(x_{a})+f(P,a)\cdot G^{a}(\boldsymbol{\omega})+\mathbf{E}(a)-\mathbf{T}\left(f(P,a)\right)\right)\right]\\
= & \underset{a\in A}{\sum}\mathbb{E}[\mathbf{P}(a)|\mathbf{x}=x]u_{a}(x_{a})+\mathbb{E}\left[\underset{a\in A}{\sum}\mathbf{P}(a)\left(f(\mathbf{P},a)\cdot G^{a}(\boldsymbol{\omega})+\mathbf{E}(a)-\mathbf{T}\left(f(\mathbf{P},a)\right)\right)\right],
\end{array}
\]
where the first equivalence follows from $\mathbb{E}[\mathbf{P}|\mathbf{x}=x]$
being the maximizer of the problem of the representative agent and
the equivalence in the third line is implied by the definition of
the average indirect utility. Lemma 11 $(ii)$ entails the equality
in the fourth line. The equality in the fifth line follows from distributing
the expectation.

Hence, taking the partial derivative of the $a^{th}$ component of
$V$ at $u(x)$,

\[
\frac{\partial}{\partial v_{a}}V\left(\vec{u}(x)\right)\Bigg|_{\vec{v}=\vec{u}(x)}=\mathbb{E}[\mathbf{P}(a)|\mathbf{x}=x].
\]
\end{proof}
\newpage{}

\section*{Appendix C --Identification with stochastic choice data}
\begin{assumption}
Suppose the following: $(i)$ (Observables) the econometrician observes
(can consistently estimate) the unconditional distribution corresponding
to the mean generated SDSC data

\[
\left\{ \mathbb{P}(a|\mathbf{x}=x)\right\} _{a\in A},
\]
for each $x\in Supp(\mathbf{x})$, where $\mathbb{P}(a|\mathbf{x}=x)=\mathbb{E}[\mathbf{P}(a)|\mathbf{x}=x]$
and the expectation is over the joint distribution of $(\boldsymbol{\omega},\mathbf{E},\mathbf{T})$;
and $(ii)$ (Full support) for each $x\in Supp(\mathbf{x})$, $\mathbb{E}[\mathbf{P}(a)|\mathbf{x}=x]>0$
for each $a\in A$.
\end{assumption}

\subsection*{Identification of good-specific utility indices}
\begin{defn}
The points $\frac{\partial}{\partial x_{a,p}}u_{a}(x_{a})|{}_{x_{a}=x_{a}^{*}}$
and $\frac{\partial}{\partial x_{b,q}}u_{b}(x_{b})|_{x_{b}=x_{b}^{*}}$
are said to be paired if they exist and the following conditions hold:
$(i)$ there exists a known value $x^{*}\in Supp(\mathbf{x})$ whose
$a^{th}$ andd $b^{th}$ components are $x_{a}^{*}$ and $x_{b}^{*}$,
respectively; $\frac{\partial}{\partial x_{a,p}}\mathbb{E}[\mathbf{P}(a)|\mathbf{x}=x]|_{x_{a}=x_{a}^{*}}$
and $\frac{\partial}{\partial x_{b,q}}\mathbb{E}[\mathbf{P}(b)|\mathbf{x}=x]|_{x_{b}=x_{b}^{*}}$
exist; $(iii)$ $V$ is twice continuously differentiable in a neighbourhood
of $u(x^{*})$; and $\frac{\partial^{2}}{\partial v_{a,}\partial v_{b}}V\left(v\right)|_{v=u(x^{*})}\neq0$.
If, in addition, $\frac{\partial}{\partial x_{a,p}}u_{a}(x_{a})|_{x_{a}=x_{a}^{*}}\neq0$
and $\frac{\partial}{\partial x_{b,q}}u_{b}(x_{b})|_{x_{b}=x_{b}^{*}}\neq0$,
then they are said to be strictly paired.
\end{defn}
\begin{defn}
Let Assumption 6 hold and suppose $x_{a,p}$ and $x_{b,q}$ are regressors
specific to $a$ and $b$, respectively. If $\frac{\partial}{\partial x_{a,p}}u_{a}(x_{a})|_{x_{a}=x_{a}^{*}}$
and $\frac{\partial}{\partial x_{b,q}}u_{b}(x_{b})|_{x_{b}=x_{b}^{*}}$
are paired and $\frac{\partial}{\partial x_{b,q}}u_{b}(x_{b})|_{x_{b}=x_{b}^{*}}\neq0$,
then there is some known $x^{*}\in Supp(\mathbf{x})$ such that

\begin{equation}
\dfrac{\frac{\partial}{\partial x_{a,p}}\mathbb{E}[\mathbf{P}(b)|\mathbf{x}=x]\Big|_{x=x^{*}}}{\frac{\partial}{\partial x_{b,q}}\mathbb{E}[\mathbf{P}(a)|\mathbf{x}=x]\Big|_{x=x^{*}}}=\dfrac{\frac{\partial}{\partial x_{a,p}}u_{a}(x_{a})\Big|_{x_{a}=x_{a}^{*}}}{\frac{\partial}{\partial x_{b,q}}u_{b}(x_{b})\Big|_{x_{b}=x_{b}^{*}}}.\label{eq:9}
\end{equation}
Let Assumption 7 hold, then the right-hand side of (\ref{eq:9}) is
identified.
\end{defn}
\begin{lem}
Let all the assumptions in Lemma 7 hold except for Asssumption 3 and
4. Let Assumptions 6 and 7 hold. Then, it follows that $\vec{u}(x)$
is identified over $Supp(\mathbf{x})$ under the following normalization:
$(i)$ $\frac{\partial}{\partial x_{b,q}}u_{b}(x_{b})|_{x_{b}=x_{b}^{*}}\in\{-1,1\}$
for a tuple $(b,q,x_{b}^{*})$ such that $x_{b}^{*}\in Supp(\mathbf{x}_{b})$;
and $(ii)$ for each $a\in A$, $\vec{u}_{a}(0_{L_{a}})=0$, where
$0_{L_{a}}\in Supp(\mathbf{x}_{a})$. 
\end{lem}

\subsubsection*{Identification of average indirect utility}
\begin{lem}
Let conditions in Lemma 8 hold, except for Assumptions 3 and 4. Suppose
Assumptions 6 and 7 hold. Then, it follows that for $x,x'\in Supp(\mathbf{x})$,
$V\left(\vec{u}(x')\right)-V\left(\vec{u}(x)\right)$ and $\bar{D}\left(\mathbb{E}[\mathbf{P}|\mathbf{x}=x']\right)-\bar{D}\left(\mathbb{E}[\mathbf{P}|\mathbf{x}=x]\right)$
are identified. In particular,

\[
V\left(\vec{u}(x')\right)-V\left(\vec{u}(x)\right)=\intop_{0}^{1}\sum_{a\in A}\mathbb{E}[\mathbf{P}(a)|\mathbf{x}=x(t)]\left(u_{a}(x_{a}')-u_{a}(x_{a})\right)dt.
\]
\end{lem}
\begin{cor}
Let Assumptions 6 and 7 hold. Suppose $u$ is known, $V$ is finite
everywhere, and the set $u(Supp(\mathbf{x}))$ is convex. It follows
that $x,x'\in Supp(\mathbf{x})$, $V\left(u(x')\right)-V\left(u(x)\right)$
and $\bar{D}\left(\mathbb{E}[\mathbf{P}|x=x']\right)-\bar{D}\left(\mathbb{E}[\mathbf{P}|x=x]\right)$
are identified.
\end{cor}

\subsection*{Identification of counterfactual bounds}
\begin{lem}
Let Assumptions 6 and 7 hold and assume $\vec{u}$ is known. Let $x^{0}\not\in Supp(\mathbf{x})$
and assume $\mathbb{E}[\mathbf{P}|\mathbf{x}=x^{0}]$ solves

\[
\sup_{P\in\text{Conv}\mathcal{P}}\sum_{a\in A}P(a)u_{a}(x_{a}^{0})+\bar{D}(P).
\]
Then, for every integer $S$ and every sequence $x^{1},...,\text{\ensuremath{x^{S}\in Supp(\mathbf{x})}}$,

\[
\begin{array}[t]{rll}
 & \underset{a\in A}{\sum} & \mathbb{E}[\mathbf{P}(a)|\mathbf{x}=x^{0}]\left(u_{a}(x_{a}^{0})-u_{a}(x_{a}^{S-1})\right)\\
\geq & \underset{a\in A}{\sum} & \Bigg[\mathbb{E}[\mathbf{P}(a)|\mathbf{x}=x^{1}]u_{a}(x_{a}^{0})-\mathbb{E}[\mathbf{P}(a)|\mathbf{x}=x^{S-1}]u_{a}(x_{a}^{S-1})\\
 &  & -\overset{S-2}{\underset{s=1}{\sum}}\left(\mathbb{E}[\mathbf{P}(a)|\mathbf{x}=x^{s}]-\mathbb{E}[\mathbf{P}(a)|\mathbf{x}=x^{s+1}]\right)u_{a}(x^{s})\Bigg].
\end{array}
\]

\end{lem}

\end{document}